\documentclass[12pt]{article} 
\newtheorem{proposition}{Proposition}[section] 
\usepackage{color} 
\usepackage{amssymb} 
\usepackage{amsmath} 
\usepackage{graphicx}
\usepackage[font=scriptsize]{caption} 
\date{} 
\title{Statistical Inference for Mixture of Cauchy Distributions \\
\vspace{1cm}\small{{Mahdi Teimouri}}\\
Email: teimouri@aut.ac.ir\\
Department of Mathematics and Statistics, Faculty of Science and Engineering, Gonbad Kavous University, Gonbad Kavous, Iran.}
\begin{document}
\maketitle{} 
\noindent{} 
{\bf{Abstract:}}
The class of $\alpha$-stable distributions received much interest for modelling impulsive phenomena occur in engineering, economics, insurance, and physics. The lack of non-analytical form for probability density function is considered as the main obstacle to modelling data via the class of $\alpha$-stable distributions. As the central member of this class, the Cauchy distribution has received many applications in economics, seismology, theoretical and applied physics. We derive estimators for the parameters of the Cauchy and mixture of Cauchy distributions through the EM algorithm. Performance of the EM algorithm is demonstrated through simulations and real sets of data.\\\\
\noindent{} {\bf{Keyword:}}
\section{Introduction}
Despite the lack of non-analytical expression for the probability density function (pdf), the class of $\alpha$-stable distributions are becoming increasingly popular in such fields as biology, ecology, economics, finance, genetics, insurance, physics, physiology, and telecommunications. Details for the applications of $\alpha$-stable distributions in aforementioned fields can be found in \cite{Johnson2004}, \cite{Klebanov2006}, \cite{Nikias1995}, \cite{Samorodnitsky1994}, and \cite{Uchaikin1999}. There are only three exceptions including Gaussian, Levy, and symmetric Cauchy distributions whose pdf has analytical form. This can be regarded as a major obstacle in the way of using this class in practice. 
In our knowledge almost all of works to unravel the problem of estimating the parameters of $\alpha$-stable distributions have been limited to the case of $\alpha \neq1$, and no attempt has been made for the Cauchy family which corresponds to $\alpha=1$. The Cauchy distribution itself has many applications in a variety of such fields, for example, as seismology \cite{Kagan1992}, applied physics \cite{Stapf1996} and \cite{Min1996}, theoretical physics \cite{Lorentz1906}, electrical engineering \cite{Winterton1992}. Similar to other members of the class of $\alpha$-stable distributions, a Cauchy distribution is introduced in terms of its characteristic function (chf). The chf of a Cauchy distributions takes different forms (parameterizations). Among them, we shall refer to two important forms, namely $S_{0}$ and $S_{1}$, see \cite{Nolan1998}. Let $Y$ follow a Cauchy distribution, then the chf of $Y$, $\varphi_{Y}{(t)}=E \exp (jtY)$ is given by the following, see \cite{Nolan1998}.
\begin{align} 
\label{chf} 
\varphi_{Y}{(t)}=\left\{\begin{array}{*{20}c} 
\exp\Bigl\{-\left| \sigma t \right|[1+j\frac{2}{\pi}\beta~\mathrm{sgn}(t)\log \left| t \right|]
+j\mu t\Bigr\},~~~~\mathrm{{for~ S_{1}~ param}}\\ 
\exp\Bigl\{-\left| \sigma t \right| [1+j\frac{2}{\pi}\beta~\mathrm{sgn}(t)\log \left| \sigma t \right|]+j \mu t\Bigr\},~~\mathrm{for~ S_{0}~ param}\\ 
\end{array} \right. 
\end{align} 
where $j^{2}$=-1 and $\mathrm{sgn}(.)$ is the well-known sign function. We use the generic symbols $C_{0}(\beta,\sigma,\mu)$ and $C_{1}(\beta,\sigma,\mu)$ to indicate a Cauchy distribution in forms $S_{0}$ and $S_{1}$, respectively. The $C_{0}(\beta,\sigma,\mu)$ family has three parameters: skewness $\beta \in [-1, 1]$, scale $\sigma \in \rm I\!{R}^{+}$, and location $\mu \in \rm I\!R$. If $\beta$=0, it would be the class of symmetric Cauchy distributions. In this case, both forms $S_{0}$ and $S_{1}$ are equal. If $\beta$=1 (-1), we have the class of totally skewed to the right (left) Cauchy distributions. If random variable $X$ accounts for a standard ($\mu$=0 and $\sigma$=1) Cauchy distribution, then based on representation (\ref{chf}) we have
\begin{align} 
\label{rep1} 
Y\mathop=\limits^d\left\{\begin{array}{*{20}c} 
\sigma X + \mu,~~~~~~~~~~~~~~~~~~~~~~\mathrm{{for~ S_{0} ~param,}}\\ 
\sigma X + \mu+ \frac{2}{\pi} \beta \sigma\log \left| \sigma \right|,~~~~\mathrm{for~ S_{1}~ param.}\\ 
\end{array} \right. 
\end{align} 
Hereafter, we write ${\cal{N}}(a,b)$ to denote a Gaussian distribution with mean $a$ and variance $b$ with pdf $g(.|a,b)$. The random variable $Z$ follows ${\cal{N}}(0,1)$ and random variable $P$ comes from $C_{0}(1,1,0)$ with pdf $h(.)$. Also, the random variable $Y$ with pdf $f(.)$ follows $C_{1}(\beta,\sigma,\mu)$. It should be noted that we are implementing the EM algorithm for the $C_{1}(\beta,\sigma,\mu)$ distribution. 
\subsection{Preliminaries} 
In the following, Proposition (\ref{prop1}) gives a useful representation that plays main role for implementing the EM algorithm.
\begin{proposition}\label{prop1} Suppose $Y \sim C_{1}(\beta,\sigma,\mu)$, $Z\sim N\sim{\cal{N}}(0,1)$, and $P\sim C_{0}(1,1,0)$. Then,
\begin{align} \label{prop11}
Y\mathop=\limits^d \eta \frac{N}{Z}+ \lambda P+\delta,
\end{align} 
where $\mathop=\limits^d$ denotes the equality in distribution, $\eta=\sigma\left(1-|\beta|\right)$, $\lambda=\sigma \beta$, and $\delta=\mu+\frac{2}{\pi}\lambda \log |\lambda|$. All random variables $N$, $Z$, and $P$ are mutually independent.
\end{proposition}
The proof of Proposition \ref{prop1} is given in Appendix~A.
\subsection{The EM Algorithm}
The EM algorithm, introduced in \cite{Dempster1977}, is the most popular approach for estimating the parameters of a statistical model when we encounter missing or latent observations. Assume that $\underline{\boldsymbol{\xi}}=\bigl(\underline{\boldsymbol{\xi}}_{1},\dots,\underline{\boldsymbol{\xi}}_{n}\bigr)$ denotes the vector of complete data. 
We write $\underline{\boldsymbol{\xi}}_{i}=(y_{i},z_{i},p_{i})$ to show the $i$-th member of the vector of the complete data consists of observed value $y_{i}$, the latent observations, ${z}_{i}$ and ${v}_{i}$; for $i=1,\dots,n$, see \cite{McLachlan2008}. Let $l_c\big(\Theta;\underline{\boldsymbol{\xi}}\big)$ account for the log-likelihood function of the complete data. The aim of the EM algorithm is maximizing the conditional expectation of the log-likelihood function of the complete data, $Q\bigl(\Theta \big|\Theta^{(t)}\bigr)=E\bigl(l_c(\Theta;\underline{\boldsymbol{\xi}})\big|\underline{y}, \Theta^{(t)}\bigr)$ given the vector of observed data $\underline{y}$ and a current guess, $\Theta^{(t)}$ of the parameters vector. The EM algorithm has two steps as: the expectation step (E-step) and the maximization step (M-step). At $(t+1)$-th iteration, the E-step computes $Q\bigl(\Theta\big|\Theta^{(t)}\bigr)$ and the M-step maximizes it with respect to $\Theta$ to get $\Theta^{(t+1)}$. Both steps are repeated until convergence occurs.
\section{The EM algorithm for Cauchy distribution}
Estimating the parameters of a Cauchy distribution through the EM algorithm needs a hierarchy or stochastic representation. Representation (\ref{prop11}) admits the hierarchy
\begin{align}\label{rep2}
Y | Z=z, P=p &\sim {\cal{N}}\Bigl(\delta+\lambda p,\frac{\eta^2}{z^2}\Bigr),\nonumber\\
Z &\sim {\cal{N}}(0,1),\nonumber\\
P &\sim C_{1}(1,1,0),
\end{align}
where $Z$ and $P$ are independent. Assume that $y_{1},\dots, y_{n}$ constitute a sequence of identically and independent realizations of $C_{1}(\beta,\sigma,\mu)$. The vector of the complete data related to (\ref{rep2}) is shown by $\underline{\boldsymbol{\xi}}=\bigl(\underline{\boldsymbol{\xi}}_1,\dots,\underline{\boldsymbol{\xi}}_n\bigr)=\bigl((y_1,p_1,{z}_1),\dots,(y_n,p_n,{z}_n)\bigr)$ in which $\underline{z}=(z_1,\dots,z_n)$ and $\underline{p}=(p_1,\dots,p_n)$ are realizations of vectors of unobservable variables $\underline{Z}$ and $\underline{P}$, respectively. Based on hierarchy (\ref{rep2}), the complete-data log-likelihood function $l_{c}(\Theta)$ is given by:
\begin{equation}
\label{lcom} 
l_{c}(\Theta)=\mathrm{C}-n\log \eta-\frac{1}{2}\sum_{i=1}^{n}\Bigl(
\frac{y_{i}-\delta-\lambda p_{i}}{\eta}\Bigr)^{2}z_{i}^{2},\nonumber\\
\end{equation}
where $\mathrm{C}$ is a constant independent of the parameters vector $\Theta=(\eta,\lambda,\delta)^{T}$. The conditional expectation of log-likelihood of the complete data $Q\bigl(\Theta \big|\Theta^{(t)}\bigl)=E\bigl(l_c(\Theta; \underline{z},\underline{v})\big|\underline{y}, \Theta^{(t)}\bigr)$ is given by
\begin{align}\label{Q} 
Q\bigl(\Theta|\Theta^{(t)}\bigr)=&\mathrm{C}-n\log \eta -\frac{1}{2\eta^{2}}\sum_{i=1}^{n}\bigl(y_{i}-\delta\bigr)^{2}E\bigl(Z_{i}^{2}\big| y_{i}, \Theta^{(t)}\bigr)+\frac{\lambda}{\eta^{2}}\sum_{i=1}^{n}\bigl(y_{i}-\delta\bigr)E\bigl(Z_{i}^{2}P_{i}\big| y_{i}, \Theta^{(t)}\bigr)\nonumber\\
&-\frac{\lambda^{2}}{2\eta^{2}}\sum_{i=1}^{n}E\bigl(Z_{i}^{2}P_{i}^{2}\big| y_{i}, \Theta^{(t)}\bigr)+\sum_{i=1}^{n}E\Bigl(\log f\bigl(p_i\bigr)\Bigr)+\sum_{i=1}^{n}E\Bigl(\log g\bigl(z_i\bigl|0,1\bigr)\Bigr),
\end{align}
where $\Theta^{(t)}=\bigl(\eta^{(t)},\lambda^{(t)},\delta^{(t)}\bigr)^{T}$. Based on (\ref{Q}), the required conditional expectations are $E\bigl(Z_{i}^{2}P_{i}^{r}\big| y_{i}, \Theta^{(t)}\bigr)$; for $r=0,1,2$. We have
\begin{align}\label{ei}
E^{(t)}_{ri}=E\bigl(Z_{i}^{2}P_{i}^{r}\big| y_{i}, \Theta^{(t)}\bigr)=\frac{2}{\pi \eta f\bigl(y\big|\Theta\bigl)}\int_{\rm I\!{R}}\biggl[1+\biggl(\frac{y_{i}-\lambda p-\delta}{\eta}\biggr)^{2}\biggr]^{-2}p^{r}h(p)dp,
\end{align}
where, as mentioned before, $h(.)$ is the pdf of the class $C_{0}(1,1,0)$. Details for computing expectation in (\ref{ei}) are given in Appendix B. The steps of the EM algorithm are given by the following.
\begin{itemize}
\item {\bf{E-step}}: Given a current guess of $\Theta$, i.e., $\Theta^{(t)}$, compute $E^{(t)}_{ri}$; for $r=0,1,2$ and $i=1,\dots,n$.
\item {\bf{M-step}}: Update $\Theta^{(t)}$ as $\Theta^{(t+1)}$ by maximizing $Q\bigl(\Theta\big|\Theta^{(t)}\bigr)$ with respect to $\delta$, $\eta$, and $\lambda$. We obtain
\begin{align}\label{etahat}
\bigl(\eta^{(t+1)}\bigr)^{2}=&\frac{\sum_{i=1}^{n}\bigl(y_{i}-\delta^{(t)}\bigr)^{2}
E^{(t)}_{0i}+\bigl(\lambda^{(t)}\bigr)^{2}\sum_{i=1}^{n}E^{(t)}_{2i}-2\lambda^{(t)}\sum_{i=1}^{n}\bigr(y_{i}-\delta^{(t)}\bigl)E^{(t)}_{1i}}{n},
\end{align}
\begin{align}\label{thetahat}
\lambda^{(t+1)}=&\frac{\sum_{i=1}^{n}\bigl(y_{i}-\delta^{(t)}\bigr)
E^{(t)}_{1i}}{\sum_{i=1}^{n}E^{(t)}_{2i}},
\end{align}
and
\begin{align}\label{muhat}
\delta^{(t+1)}=&\frac{\sum_{i=1}^{n}y_{i}E^{(t)}_{0i}-\lambda^{(t+1)}E^{(t)}_{1i}}{\sum_{i=1}^{n}E^{(t)}_{0i}},
\end{align}
where finite quantities $E^{(t)}_{ri}$; for $r=0,1,2$ are approximated by the Monte Carlo method.
\end{itemize} 
If the required expectations in (\ref{etahat})-(\ref{muhat}) are evaluated exactly, then the EM algorithm converges to the global maximum for a small number of iterations. Once we have estimated the parameters $\eta$ and $\lambda$ through the EM algorithm as $\hat{\eta}$ and $\hat{\lambda}$, respectively, the parameter $\beta$ is obtained as the root of the equation $\hat{\eta}(1-|{\beta}|)-\beta{\hat{\lambda}}=0$. Having $\hat{\beta}$, we estimate $\sigma$ and $\mu$ as $\hat{\sigma}={\hat{\lambda}}/(1-|{\hat{\beta}}|)$ and $\delta-2/\pi {\hat{\beta}} {\hat{\sigma}} \log |{\hat{\beta}} {\hat{\sigma}}|$, respectively, for $|{\hat{\beta}}| \neq 1$.

\par When data come from $C_{0}(\beta,\sigma,\mu)$ distribution, the EM-based estimations of the scale and skewness parameters remain unchanged but the location parameter is estimated as $\hat{\mu}+2/\pi \hat{\beta}\hat{\sigma} \log \hat{\sigma}$.
\section{The EM algorithm for mixture of Cauchy distributions}
The mixture of $\alpha$-stable distributions has been considered as a suitable model for statistical analysis of the phenomena with multimodal and heavy-tailed relative frequency, see \cite{Casarin2004}, \cite{Gonzalez2009}, \cite{Gonzalez2010}, and \cite{Teimouri2018}. Let $\underline{y}=(y_{1},\dots,y_{n})$ denote the vector of observed values of a K-component mixture of $C_{1}(\beta,\sigma,\mu)$ distributions. The pdf of mixture model, i.e., ${\cal{F}}(y|\underline{\Theta})$ is represented as 
\begin{align}\label{mix}
{\cal{F}}(y|\underline{\Theta})=\sum_{j=1}^{K}\omega_{j}f(y|\underline{\Theta}_{j}),
\end{align}
where K denotes the number of components, $\underline{\Theta}_{j}=\bigl(\eta_{j},\lambda_{j},\delta_{j})$; for $j=1,\dots,K$, $f(.|\underline{\Theta}_{j})$ is pdf in class $C_{0}\bigl(\beta_{j},\sigma_{j},\mu_{j}\bigr)$, $\underline{\Theta}=\bigl(\underline{\Theta}_{1},\dots,\underline{\Theta}_{K}\bigr)$, and $w_{j}$s; for $j=1,\dots,K$ are non-negative values that sum to one. 
The complete data related to the mixture model given in (\ref{mix}) is shown by $\underline{\boldsymbol{\xi}}=\left(\underline{\boldsymbol{\xi}}_1,\dots,\underline{\boldsymbol{\xi}}_n\right)=\bigl((y_{1},p_{1},z_{1},\underline{\boldsymbol{b}}_{1}),\dots,(y_{n},p_{n},z_{n},\underline{\boldsymbol{b}}_{n})\bigr)$ in which $p_{1},\dots, p_{n}$ and $z_{1},\dots, z_{n}$ are realizations of unobservable variables $P_{1},\dots, P_{1}$ and $Z_{1},\dots, Z_{1}$; also $\underline{\boldsymbol{b}}=(\underline{\boldsymbol{b}}_1,\dots, \underline{\boldsymbol{b}}_n)$ in which $\underline{\boldsymbol{b}}_i=(b_{i1},\dots, b_{iK})$ is realizations of the latent vector $\underline{\boldsymbol{B}}_{i}=\bigl(B_{i1},\dots, B_{iK}\bigr)$. For each observed value such as $y_{i}$; for $i=1,\dots, n$, one of the components of $\underline{\boldsymbol{B}}_{i}$ is one and others are zero. For instance, if $y_{i}$ comes from the $j$-th component, then $B_{ij}=1$ and $B_{ik}=0$; for $k=1,\dots,K$ and $k \neq j$. If random variable $Y$ has a pdf of the form ${\cal{F}}(y|\underline{\Theta})$, the $Y$ admits the hierarchy given by the following.
\begin{align}\label{rep}
Y_{i}\big| Z_{i}=z_{i}, P_{i}=p_{i}, {B}_{ij}=1&\sim {\cal{N}}\Biggl(\delta_{j}+\lambda_{j}p_{i},\frac{\eta_{j}^{2}}
{z^{2}_{i}}\Biggr),\nonumber\\
P_{i}&\sim C_{0}(1,1,0),\nonumber\\
Z_{i}&\sim {\cal{N}}(0,1),\nonumber\\
{B}_{i}& \sim {\cal{M}}ultinomial(1,\omega_{1},\dots,\omega_{K}),
\end{align}
for $j=1,\dots, K$ and $i=1,\dots,n$. 
Suppose that K is known, based on representation (\ref{rep}), it is not hard to check that the log-likelihood function of the complete data is
\begin{align}
l_{c}\bigl(\underline{\Theta};\underline{p}, \underline{z},\underline{\boldsymbol{b}}\bigr)=\mathrm{C}+\sum_{i=1}^{n}\sum_{j=1}^{K} b_{ij} \log \omega_{j}-\sum_{i=1}^{n}\sum_{j=1}^{K} b_{ij} \Biggl(\log \eta_{j}+\frac{\bigl(y_i-\delta_{ j}-p_{i}\lambda_{j}\bigr)^2 z^{2}_{i}}{2 \eta_{j}^{2}}\Biggr),\nonumber
\end{align} 
where $\mathrm{C}$ is a constant independent of the parameters vector $\underline{\Theta}=\bigl(\underline{\Theta}_{1},\dots, \underline{\Theta}_{K}\bigr)=\bigl((\delta_{1},\eta_{1},\lambda_{1})$\\
$,\dots,(\delta_{K},\eta_{K},\lambda_{K})\bigr)$. After simplifications, the conditional expectation of the log-likelihood function of complete data $Q\bigl(\underline{\Theta}\big |\underline{\Theta}^{(t)}\bigr)=E\bigl(l_c(\underline{\Theta};\underline{p}, \underline{z},\underline{b})\big |\underline{y}, \underline{\Theta}^{(t)}\bigr)$, at the $t$-th iteration of the EM algorithm is
\begin{align}
Q\bigl(\underline{\Theta}|\underline{\Theta}^{(t)}\bigr)=&\sum_{i=1}^{n}\sum_{j=1}^{K} E^{(t)}_{1ij} \log \omega_{j}-\sum_{i=1}^{n}\sum_{j=1}^{K} E^{(t)}_{1ij} \log \eta_{j}\nonumber\\
&-\frac{1}{2}\sum_{i=1}^{n}\sum_{j=1}^{K} \biggl(\frac{y_i-\delta_{ j}}{\eta_{j}} \biggr)^{2} E^{(t)}_{1ij} E^{(t)}_{2ij}+\sum_{i=1}^{n}\sum_{j=1}^{K} \frac{\bigl(y_i-\delta_{ j} \bigr)\lambda_{ j}}{\eta_{j}^{2}} E^{(t)}_{1ij} E^{(t)}_{3ij}\nonumber\\
&-\frac{1}{2}\sum_{i=1}^{n}\sum_{j=1}^{K} \biggl(\frac{\lambda_{ j}}{\eta_{j}} \biggr)^{2} E^{(t)}_{1ij} E^{(t)}_{4ij},\nonumber
\end{align}
where
\begin{align}
{{E}^{(t)}_{1ij}}=&E\left(B_{ij}\Big |\underline{\Theta}^{(t)}, y_{i}\right)=\frac{{\pi}^{(t)}_{j}f\big(y_{i}\big|\underline{\Theta}^{(t)}_{j}\big)}{\sum_{j=1}^{K}\pi^{(t)}_{j}f\big(y_{i}\big|\underline{\Theta}^{(t)}_{j}\big)},\label{E1ij}
\\
{{E}^{(t)}_{2ij}}=&E\Big(B_{ij}Z^{2}_{i}\Big |\underline{\Theta}^{(t)}, y_{i}\Big)=
{{E}^{(t)}_{1ij}}E\Big(Z^{2}_{i}\Big |\underline{\Theta}^{(t)}_{j}, y_{i}\Big),\label{E2ij}
\\
{{E}^{(t)}_{3ij}}=&E\Big(B_{ij}Z^{2}_{i}P_{i}\Big |\underline{\Theta}^{(t)}, y_{i}\Big)=
{{E}^{(t)}_{1ij}}E\Big(Z^{2}_{i}P_{i}\Big |\underline{\Theta}^{(t)}_{j}, y_{i}\Big),\label{E3ij}
\\
{{E}^{(t)}_{4ij}}=&E\Big(B_{ij}Z^{2}_{i}P^{2}_{i}\Big |\underline{\Theta}^{(t)}, y_{i}\Big)
={{E}^{(t)}_{1ij}}E\Big(Z^{2}_{i}P_{i}^{2}\Big |\underline{\Theta}^{(t)}_{j}, y_{i}\Big), \label{E4ij}
\end{align}
in which $\underline{\Theta}^{(t)}_{j}=\Big(\delta_{j}^{(t)}$, $\eta_{j}^{(t)}$, $\lambda_{j}^{(t)}\Big)$. 
All quantities of the form $E\big(Z^{2}_{i}P^{r}\big |\underline{\Theta}^{(t)}_{j}, y_{i}\big)$; for $j=1,\dots, K$ and $i=1,\dots,n$, appeared in (\ref{E2ij})-(\ref{E4ij}) can be computed using the method described in Appendix B. It follows that
\begin{align}
E\Big(Z^{2}_{i}P^{r}_{i}\Big |\underline{\Theta}^{(t)}_{j}, y_{i}\Big)=\frac{2}{\pi \eta^{(t)}_{j}f\bigl(y_{i}\big|\underline{\Theta}^{(t)}_{j}\bigl)}\int_{{\rm I\!R}}\frac{u^{r}h(u)du}{\Biggl[1+\Bigl(\frac{y_{i}-\delta^{(t)}_{j}-\lambda^{(t)}_{j}{u}}{\eta^{(t)}_{j}}\Bigr)^{2}\Biggr]^{2}},
\nonumber
\end{align}
where $r=0,1,2$. The quantities $E\big(Z^{2}_{i}P^{r}_{i}\big |\underline{\Theta}^{(t)}_{j}, y_{i}\big)$; for $r=0,1,2$ given in (\ref{E1ij})-(\ref{E4ij}) are approximated by Monte Carlo method similar to the method described in Appendix~B. 
The steps of the EM algorithm for the mixture of Cauchy distributions are given by the following.
\begin{itemize}
\item {\bf{E-step}}: At the $t$-th iteration, given a guess of $\underline{\Theta}$, i.e., $\underline{\Theta}^{(t)}$, the quantities $E\big(Z^{2}_{i}P^{r}_{i}\big |\underline{\Theta}^{(t)}_{j}, y_{i}\big)$; for $r=0,1,2$, $j=1,\dots,K$, and $i=1,\dots,n$, given in (\ref{E1ij})-(\ref{E4ij}) are computed.
\item {\bf{M-step}}: At the $t$-th iteration, the M step maximizes $Q\bigl(\underline{\Theta}\big|\underline{\Theta}^{(t)}\bigr)$ with respect to $\underline{\Theta}_{j}$ to obtain $\underline{\Theta}^{(t+1)}_{j}$ as
\begin{align}
{{\omega}^{(t+1)}_{j}}=&\frac{{\omega}^{(t)}_{j}f\big(y_{i}\big|\underline{\Theta}^{(t)}_{j}\big)}{\sum_{j=1}^{K}\omega^{(t)}_{j}f\big(y_{i}\big|\underline{\Theta}^{(t)}_{j}\big)},
\nonumber\\
\delta^{(t+1)}_{j}=&\frac{
\sum_{i=1}^{n}y_{i}E^{(t)}_{1ij}E^{(t)}_{2ij}
-\lambda^{(t)}_{j}\sum_{i=1}^{n}E^{(t)}_{1ij}E^{(t)}_{3ij}
}{\sum_{i=1}^{n}E^{(t)}_{1ij}E^{(t)}_{2ij}},
\nonumber\\
\Bigl(\eta^{(t+1)}_{j}\Bigr)^{2}=&\frac{
\sum_{i=1}^{n}E^{(t)}_{1ij}E^{(t)}_{2ij}\Bigl(y_i-\delta^{(t+1)}_{ j} \Bigr)^{2}}{\sum_{i=1}^{n}E^{(t)}_{1ij}}+\frac{\Bigl(\lambda^{(t)}_{j}\Bigl)^{2}\sum_{i=1}^{n}E^{(t)}_{1ij}E^{(t)}_{2ij}}{\sum_{i=1}^{n}E^{(t)}_{1ij}}\nonumber\\
&-2\lambda^{(t)}_{j}\frac{\sum_{i=1}^{n}E^{(t)}_{1ij}E^{(t)}_{3ij}\Bigl(y_i-\delta^{(t)}_{ j}\Bigl)}
{\sum_{i=1}^{n}E^{(t)}_{1ij}},\nonumber\\
\lambda^{(t+1)}_{j}=&\frac{\sum_{i=1}^{n}E^{(t)}_{1ij}E^{(t)}_{3ij}\Bigl(y_i-\delta^{(t+1)}_{ j}\Bigl)}{\sum_{i=1}^{n}E^{(t)}_{1ij}E^{(t)}_{4ij}},\nonumber
\end{align}
for $j=1,\dots,K$. 
\end{itemize}
\section{Performance analysis}
This section has three parts. Firstly, we perform a simulation study to compare the performance of the EM and ML approaches for estimating the parameters of $C_{1}(\beta,\sigma,\mu)$ distribution. 
Secondly, we carry out a simulation study to investigate the performance of the EM algorithm for estimating the parameters of a two-component mixture of Cauchy distributions.
Finally, performance of the EM algorithm is demonstrated via two sets of real data.
\subsection{Model validation via simulation}
\begin{figure}
\resizebox{\textwidth}{!}
{\begin{tabular}{ccc}
\includegraphics[width=40mm,height=40mm]{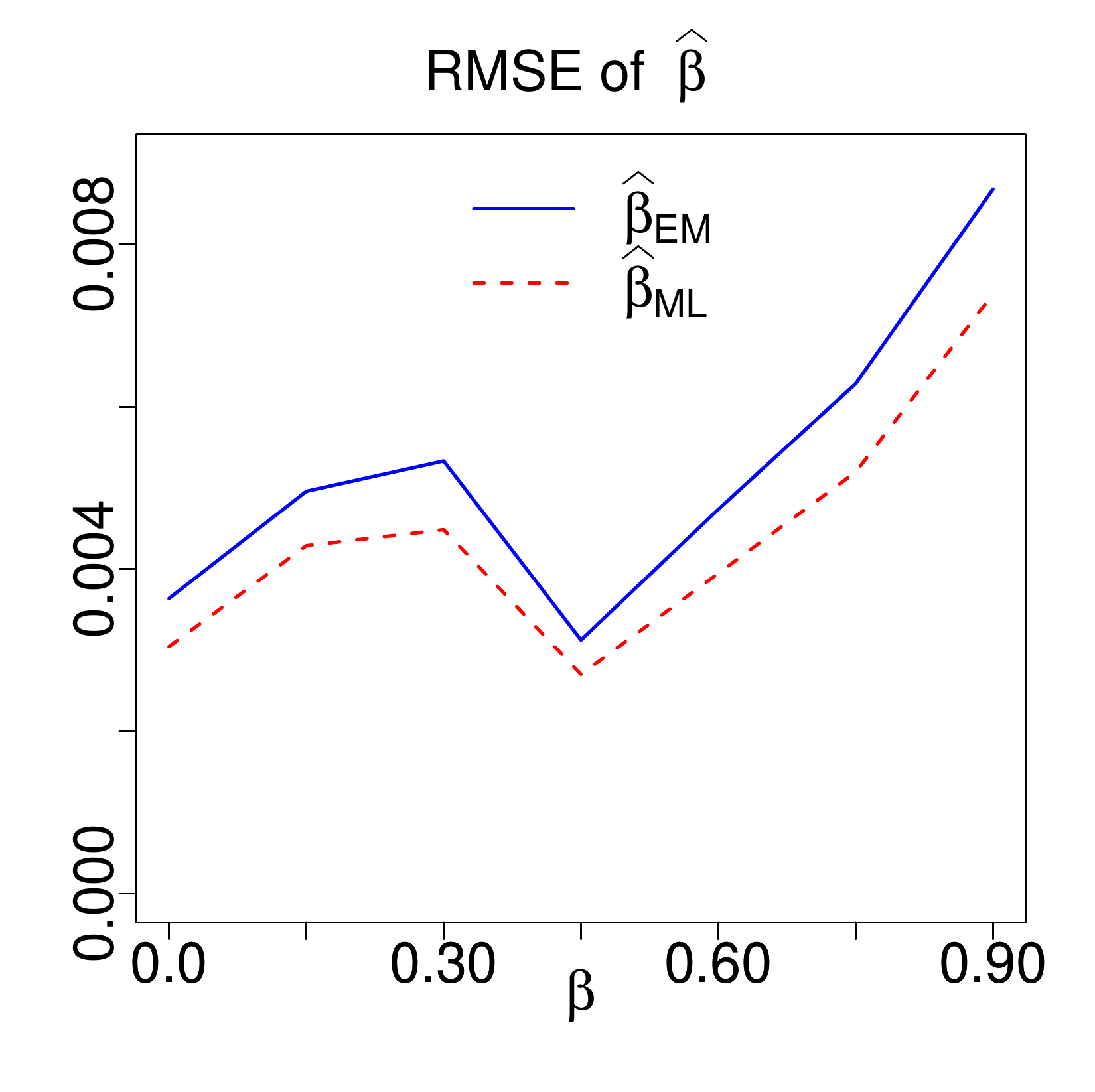}&
\includegraphics[width=40mm,height=40mm]{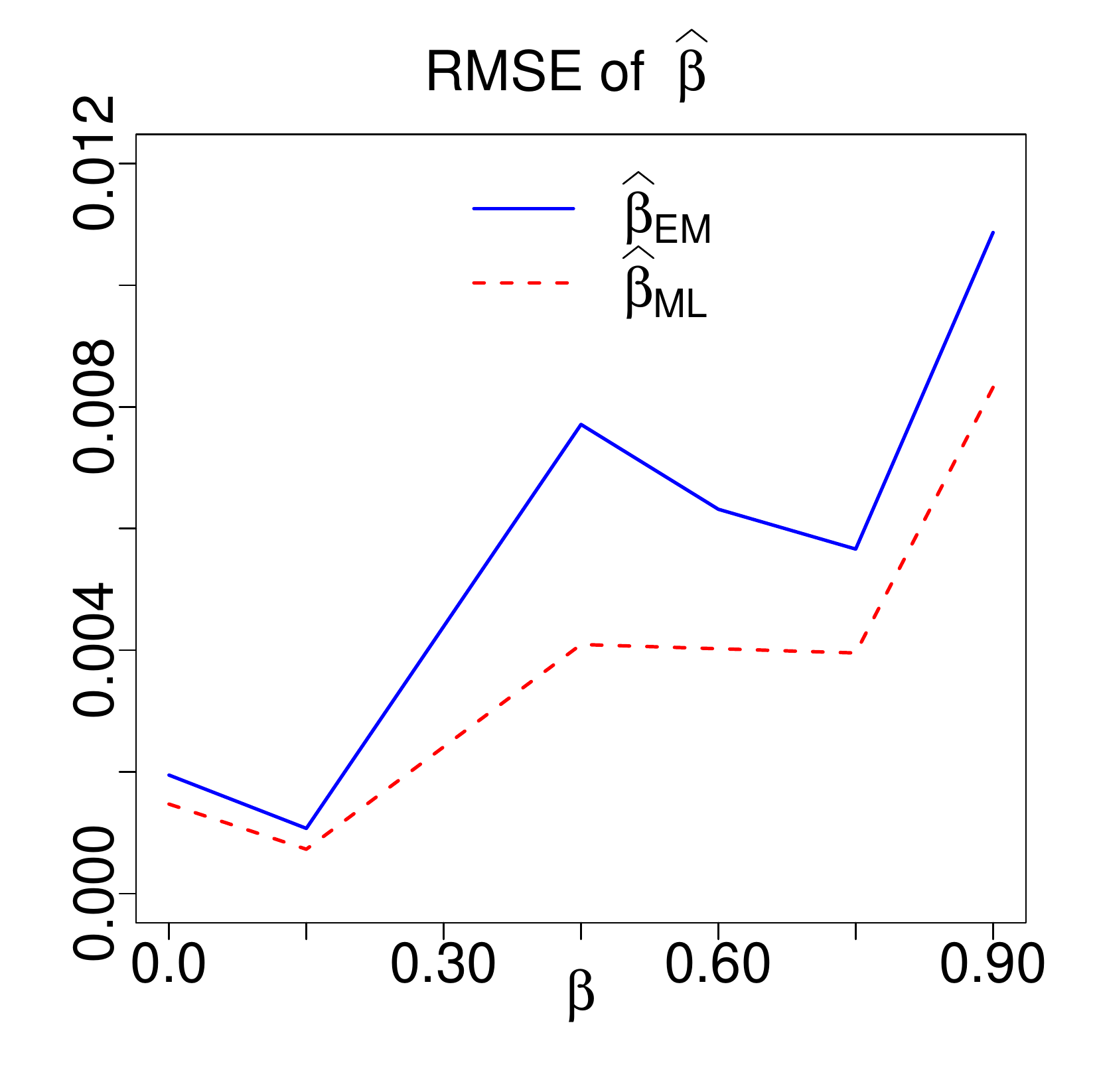}&
\includegraphics[width=40mm,height=40mm]{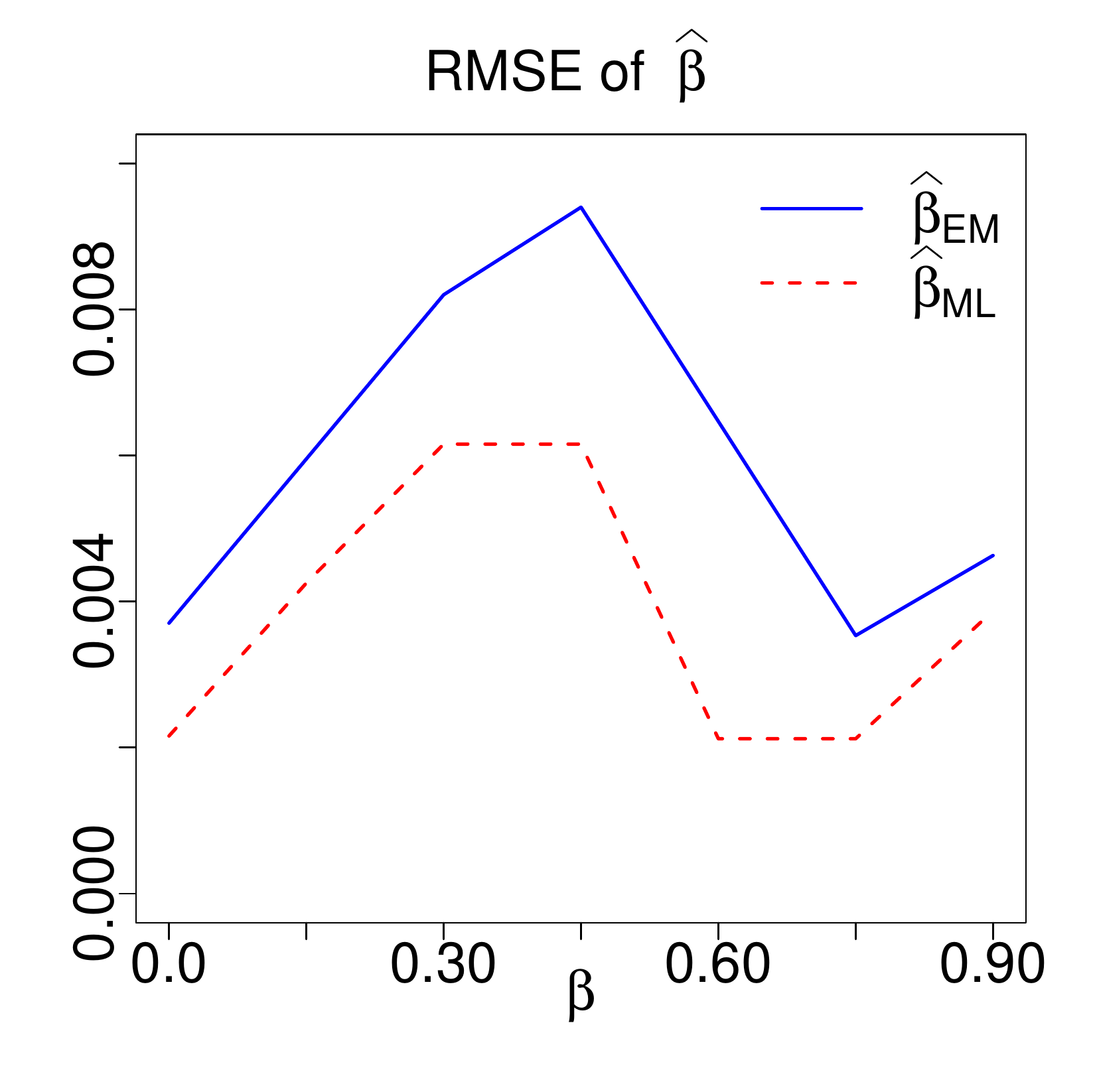}\\
\includegraphics[width=40mm,height=40mm]{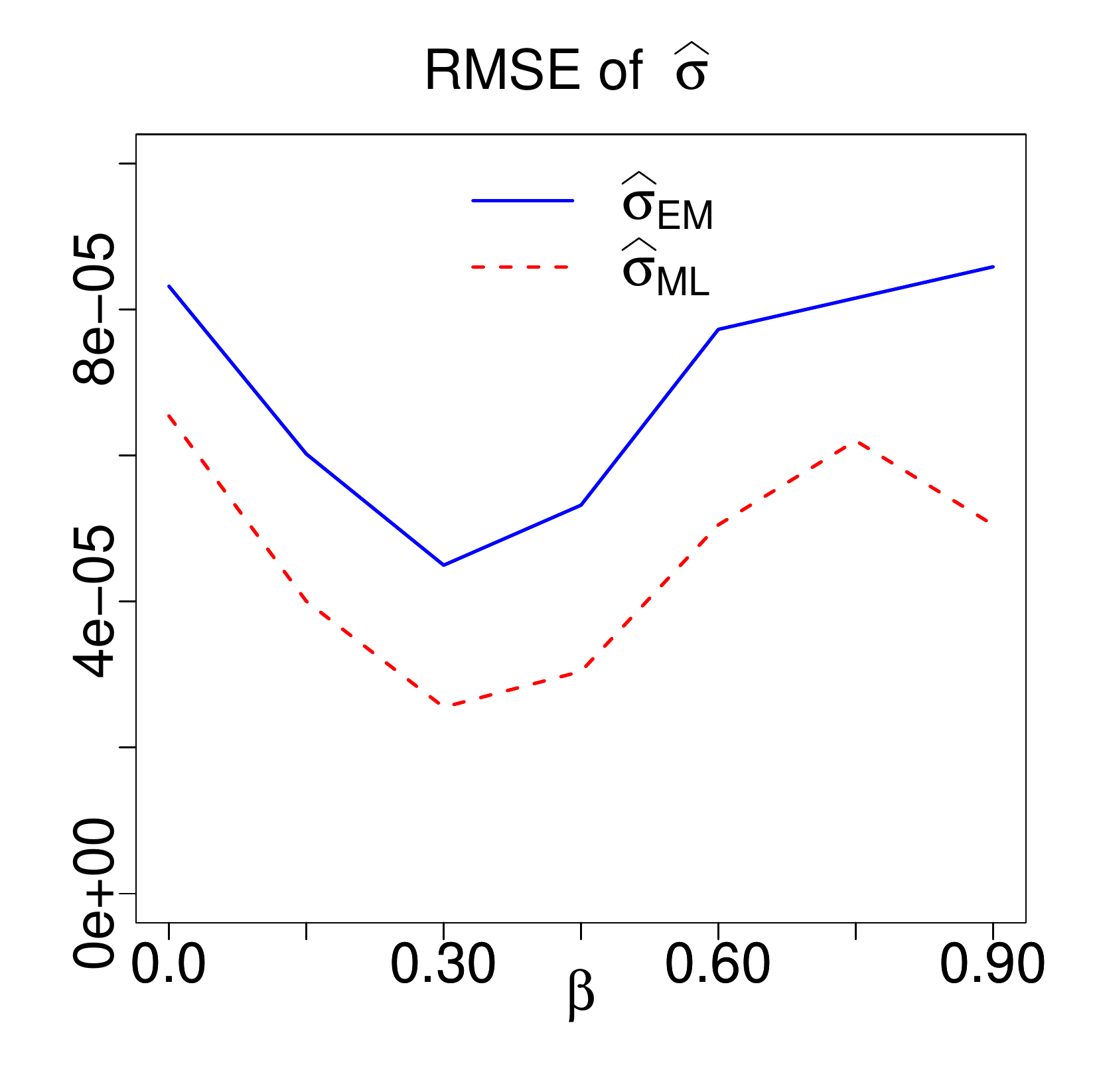}&
\includegraphics[width=40mm,height=40mm]{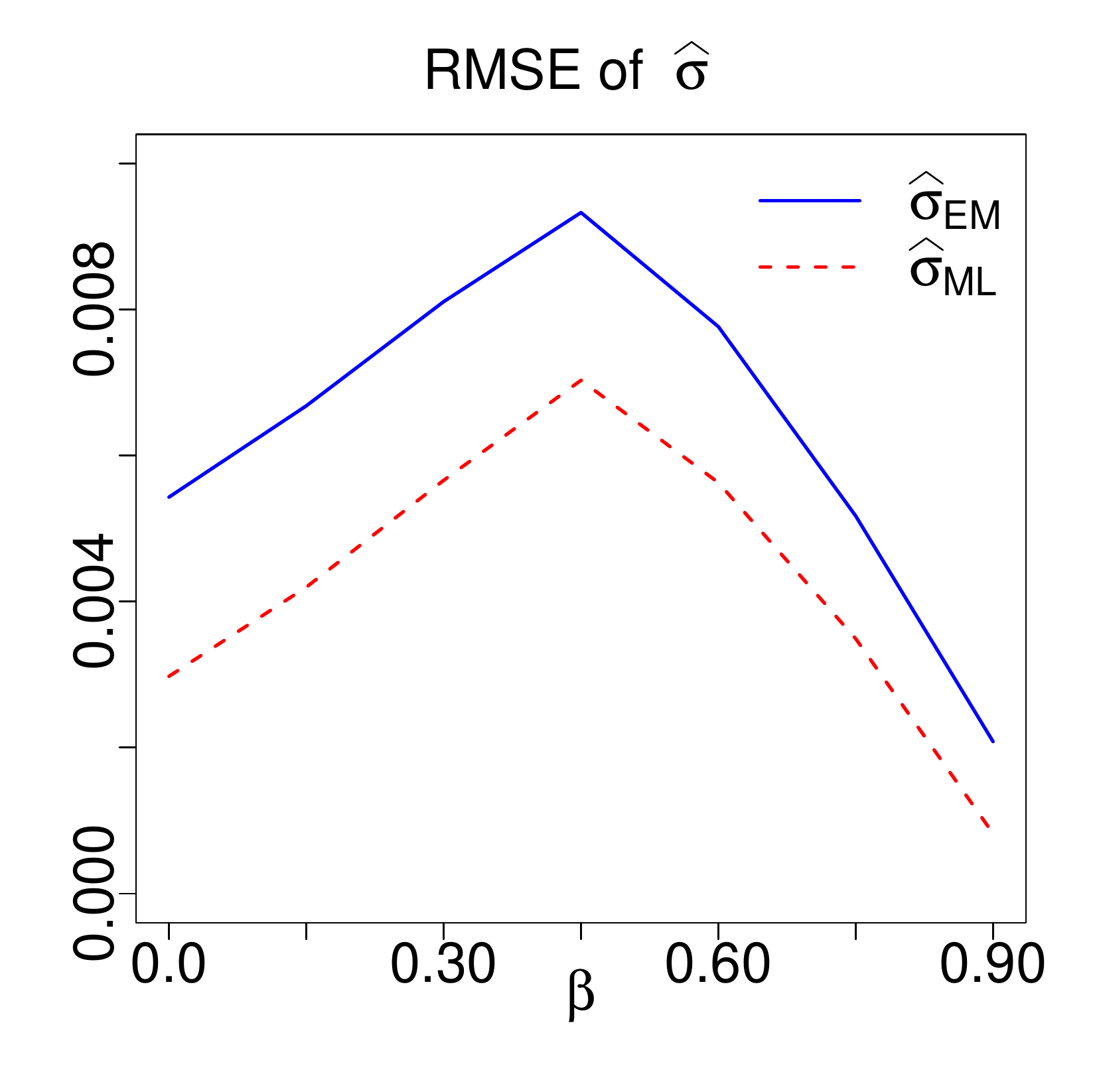}&
\includegraphics[width=40mm,height=40mm]{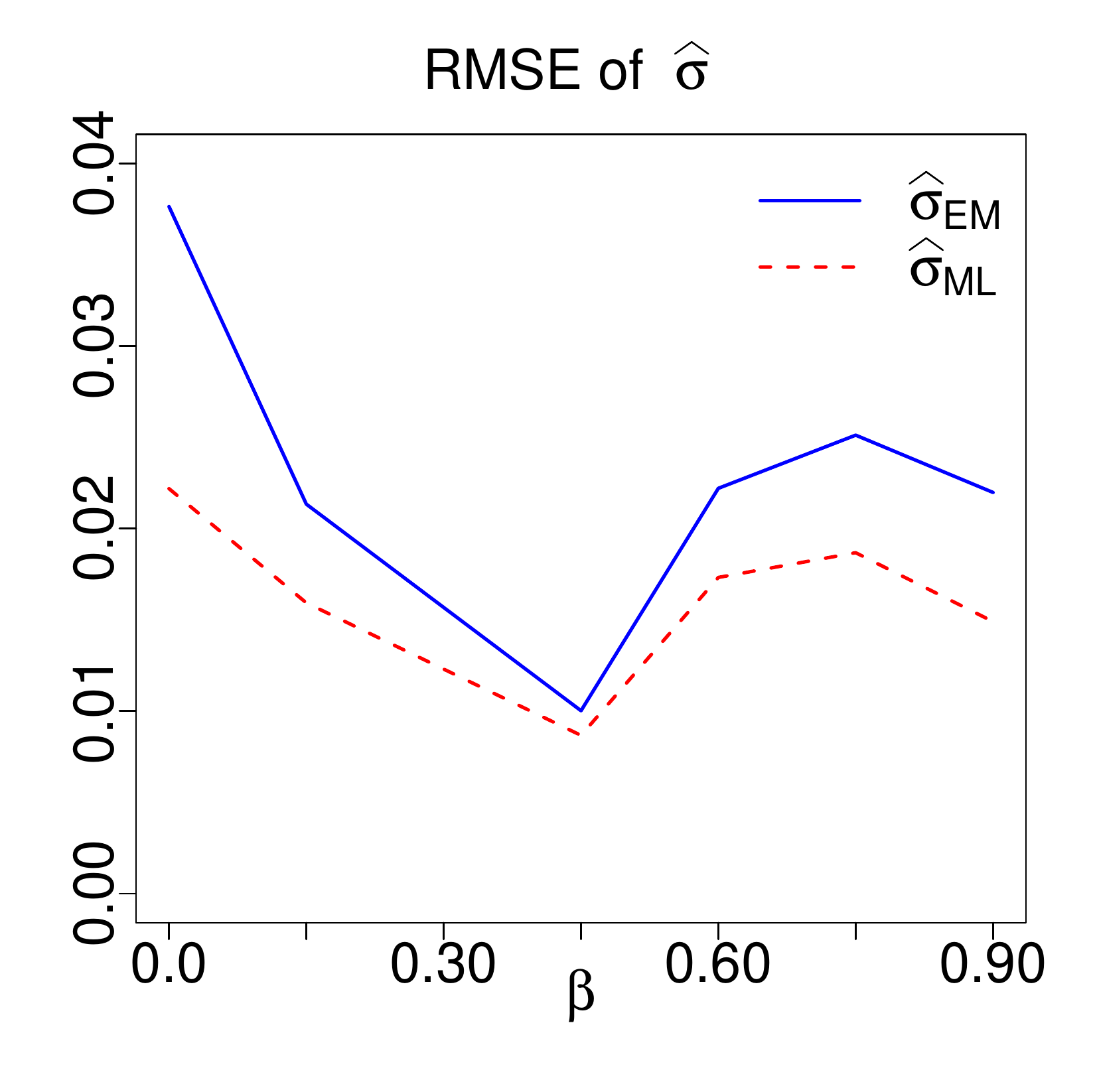}\\
\includegraphics[width=40mm,height=40mm]{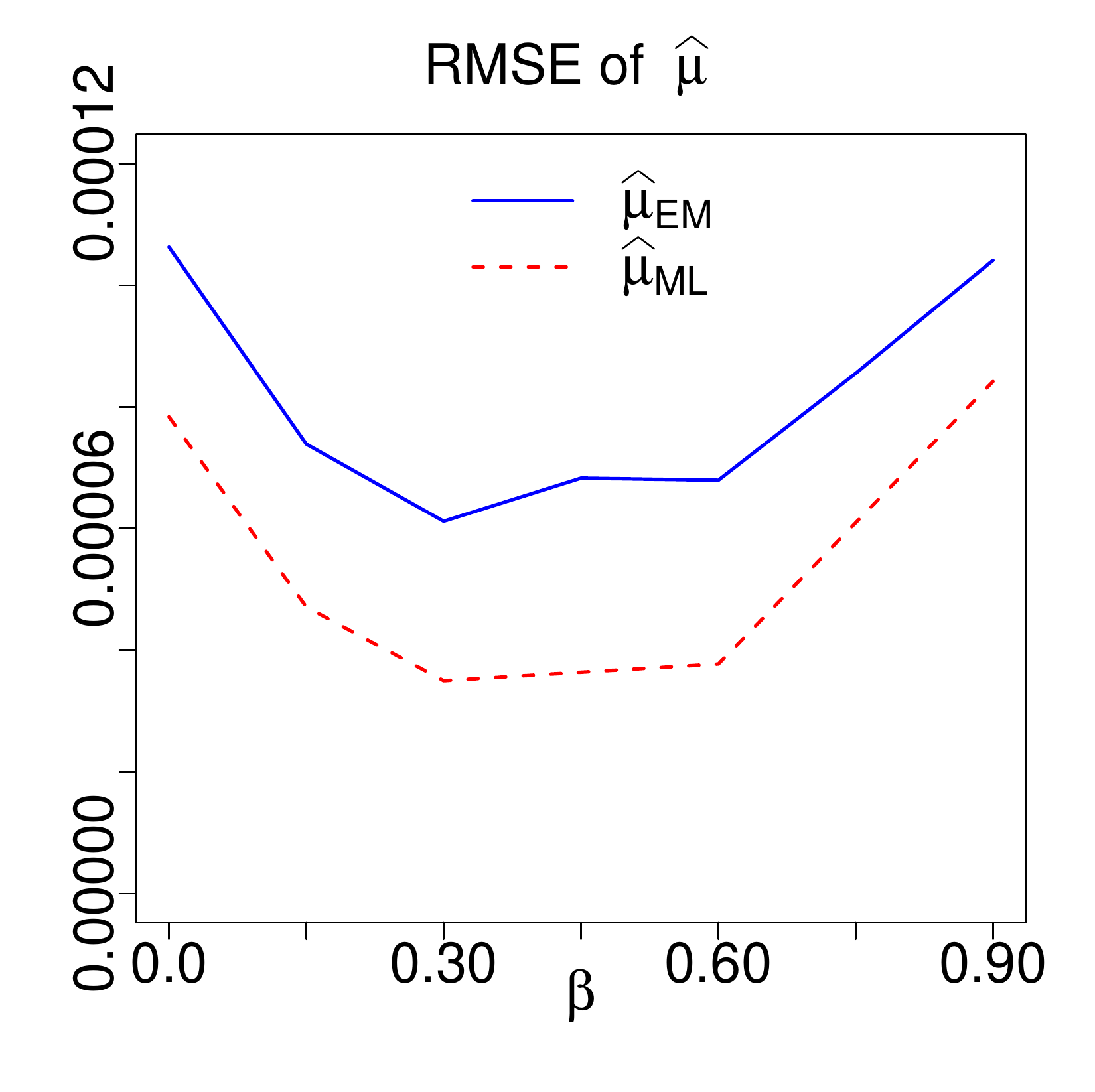}&
\includegraphics[width=40mm,height=40mm]{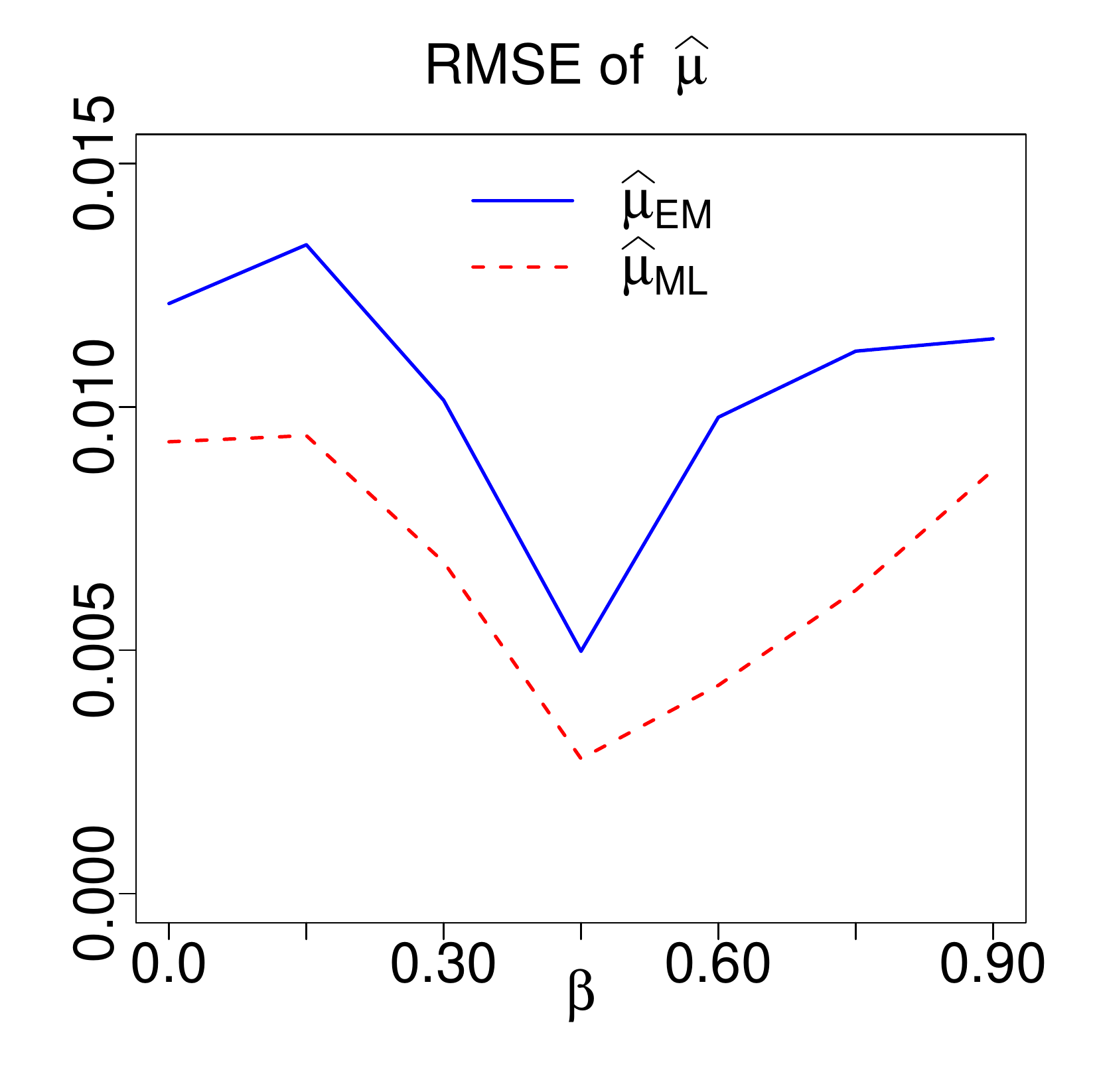}&
\includegraphics[width=40mm,height=40mm]{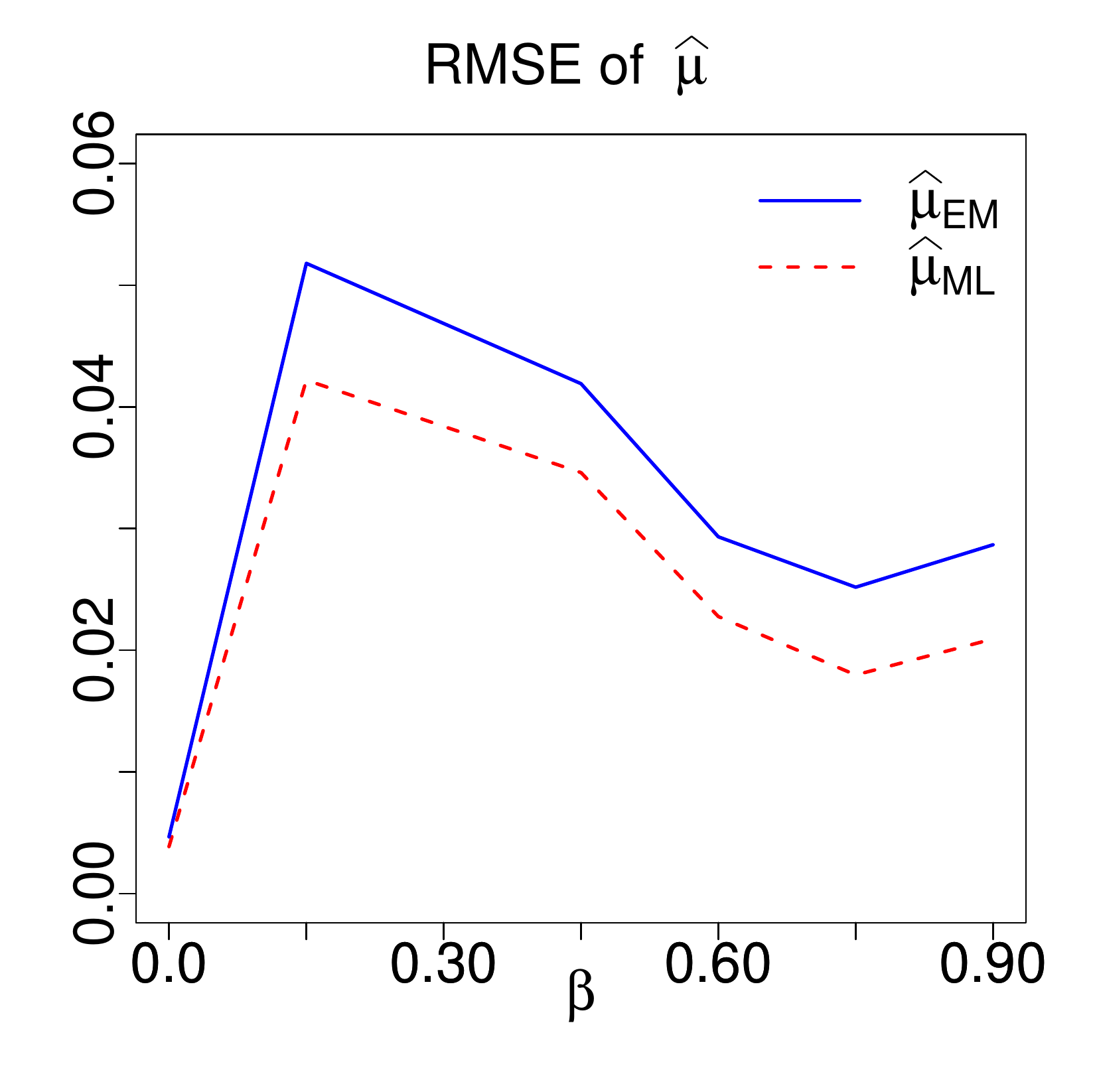}\\
\end{tabular}}
\caption{The RMSE of estimators obtained through the EM and ML approaches. The estimators are obtained when 300 realizations are generated from Cauchy distribution. In each sub-figure, the subscripts ML and EM indicate that the estimators $\hat{\beta}$, $\hat{\sigma}$, and $\hat{\mu}$ are obtained using the EM algorithm (blue solid line) or the ML approach (red dashed line). Note that, the levels of the skewness parameter on the horizontal axis are 0.0,0.15,0.30,0.45,0.60,0.75, and 0.90. The sub-figures in the first, second, and the third columns correspond to $\sigma=0.10$, $\sigma=2$, and $\sigma=5$, respectively.}
\label{fig1}
\end{figure}

Here, we perform a simulation study to compare the performance of the EM algorithm and ML approaches for estimating the parameters of $C_{1}(\beta,\sigma,\mu)$ distribution. For this, we set $\mu=0$, $\sigma=0.1,2,5$, and $\beta=0.0,0.15,0.30,0.45,0.60,0.75,0.90$. For each setting, a sample of 300 realizations are generated from $C_{1}(\beta,\sigma,\mu)$ distribution. Comparisons between the EM and ML approaches are made based on the root of mean square error (RMSE). As Figure \ref{fig1} shows, the EM algorithm works satisfactorily and can be considered as a good competitor for the ML approach.
\subsection{EM algorithm for mixture of Cauchy distributions 
}
\begin{figure}
\resizebox{\textwidth}{!}
{\begin{tabular}{cc}
\includegraphics[width=40mm,height=25mm]{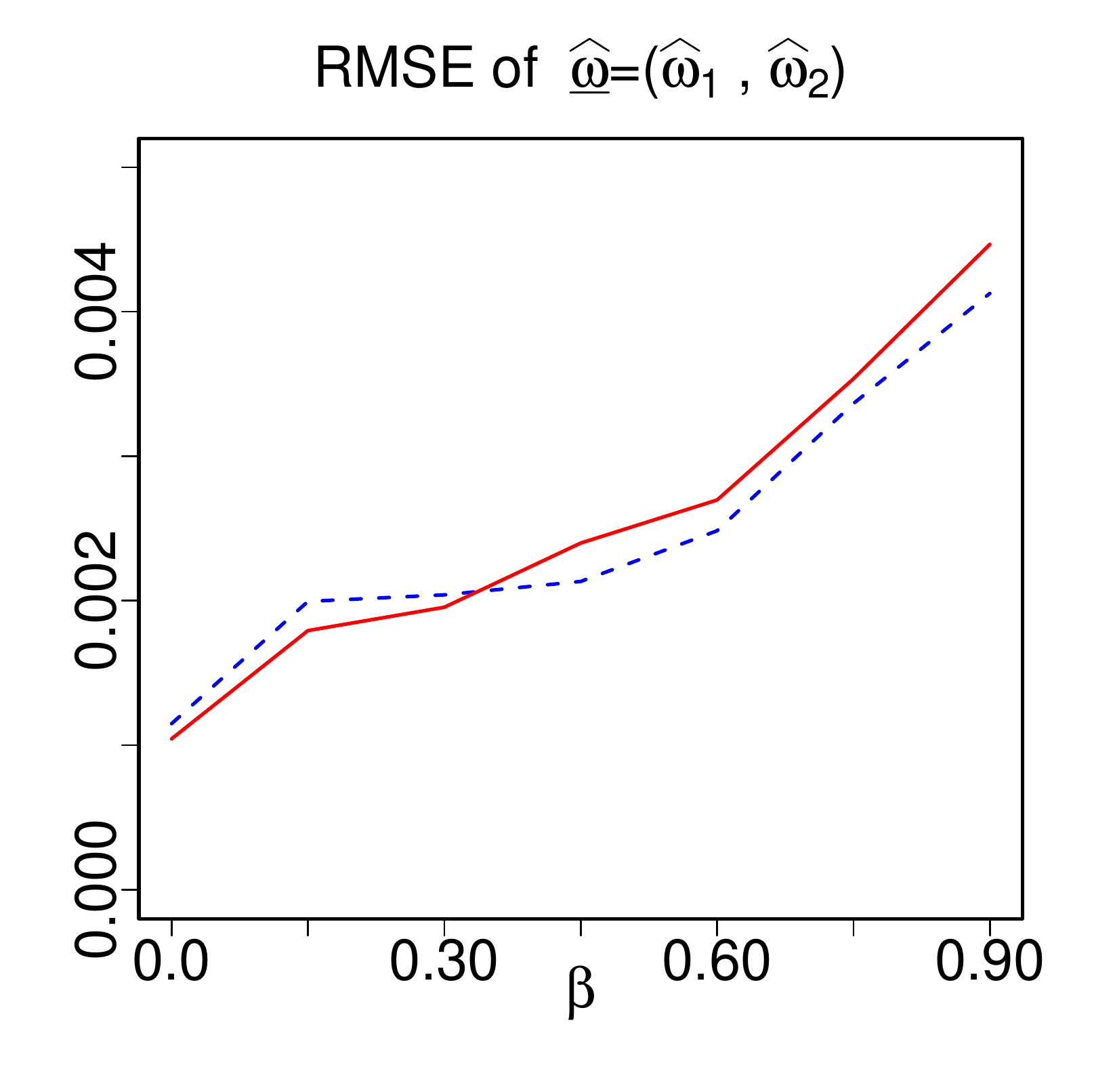}&
\includegraphics[width=40mm,height=25mm]{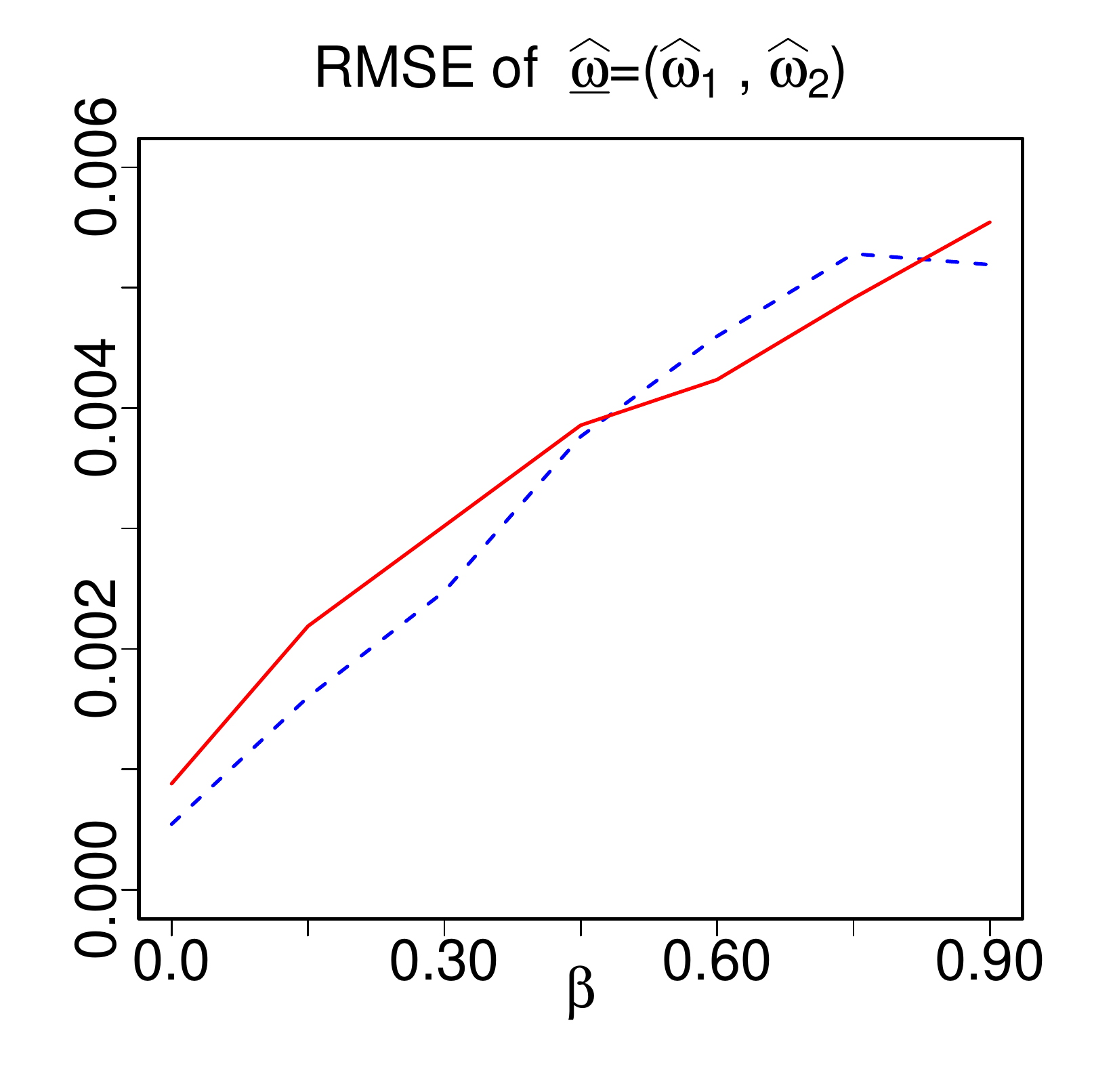}\\
\includegraphics[width=40mm,height=25mm]{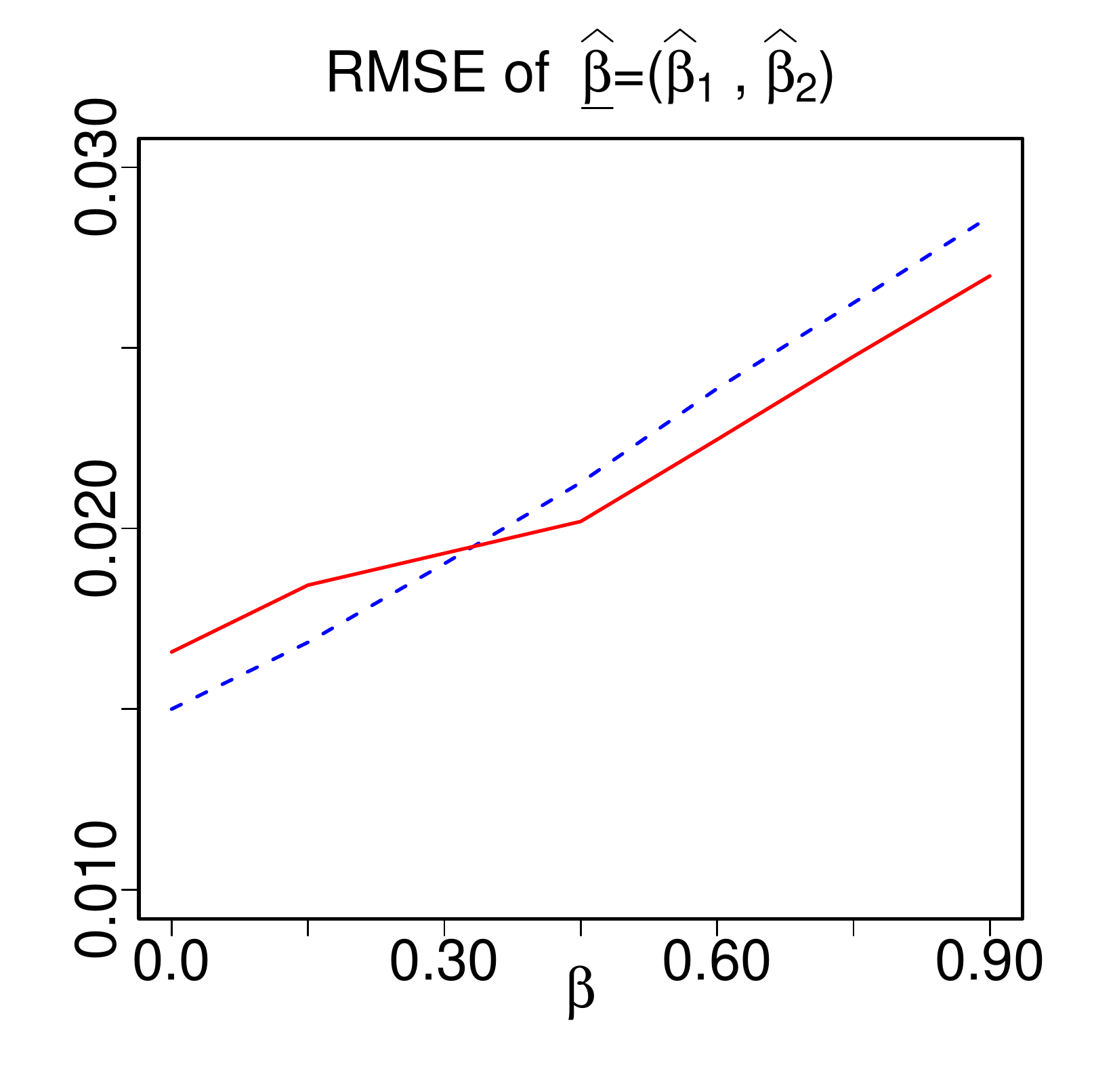}&
\includegraphics[width=40mm,height=25mm]{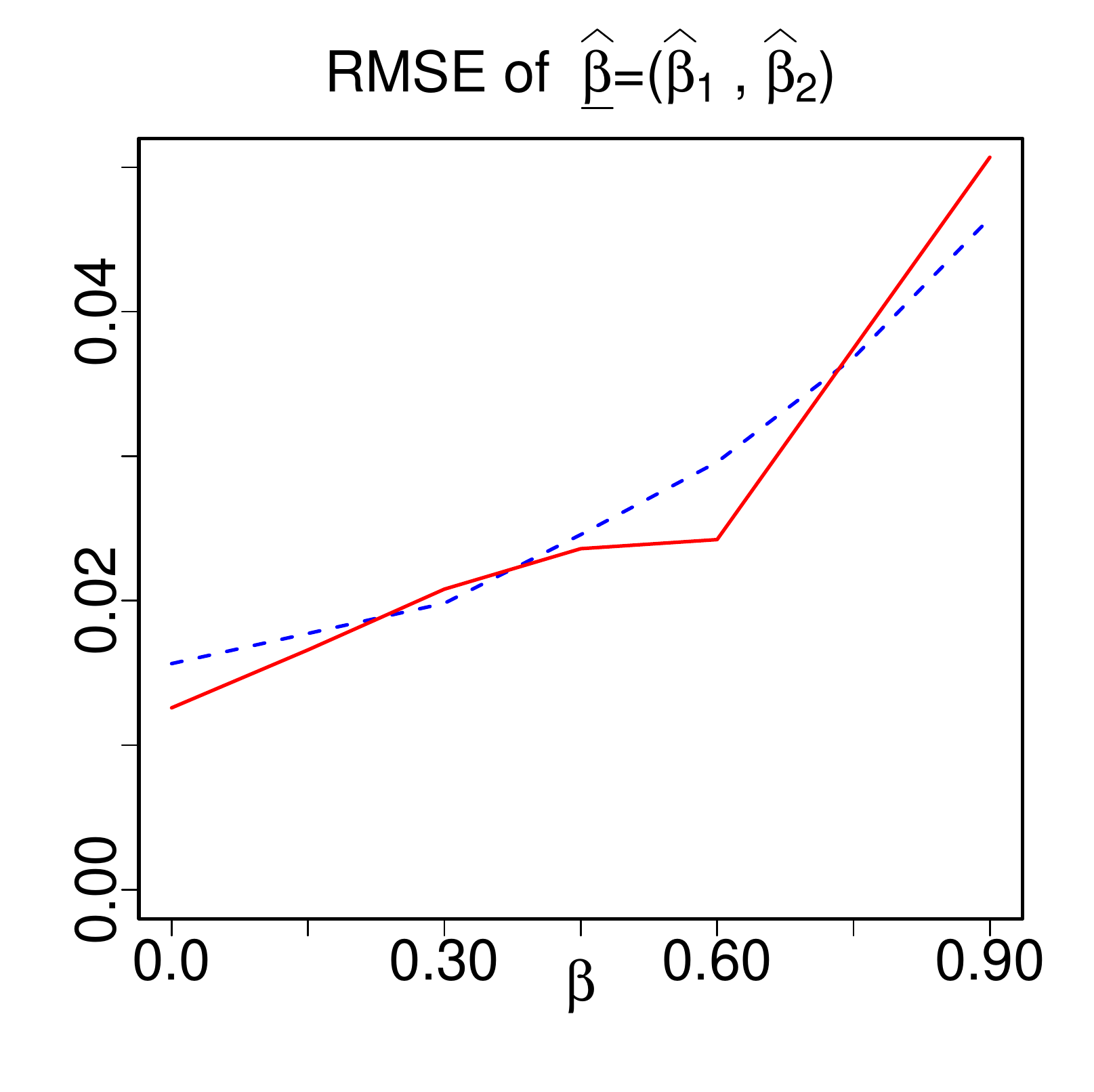}\\
\includegraphics[width=40mm,height=25mm]{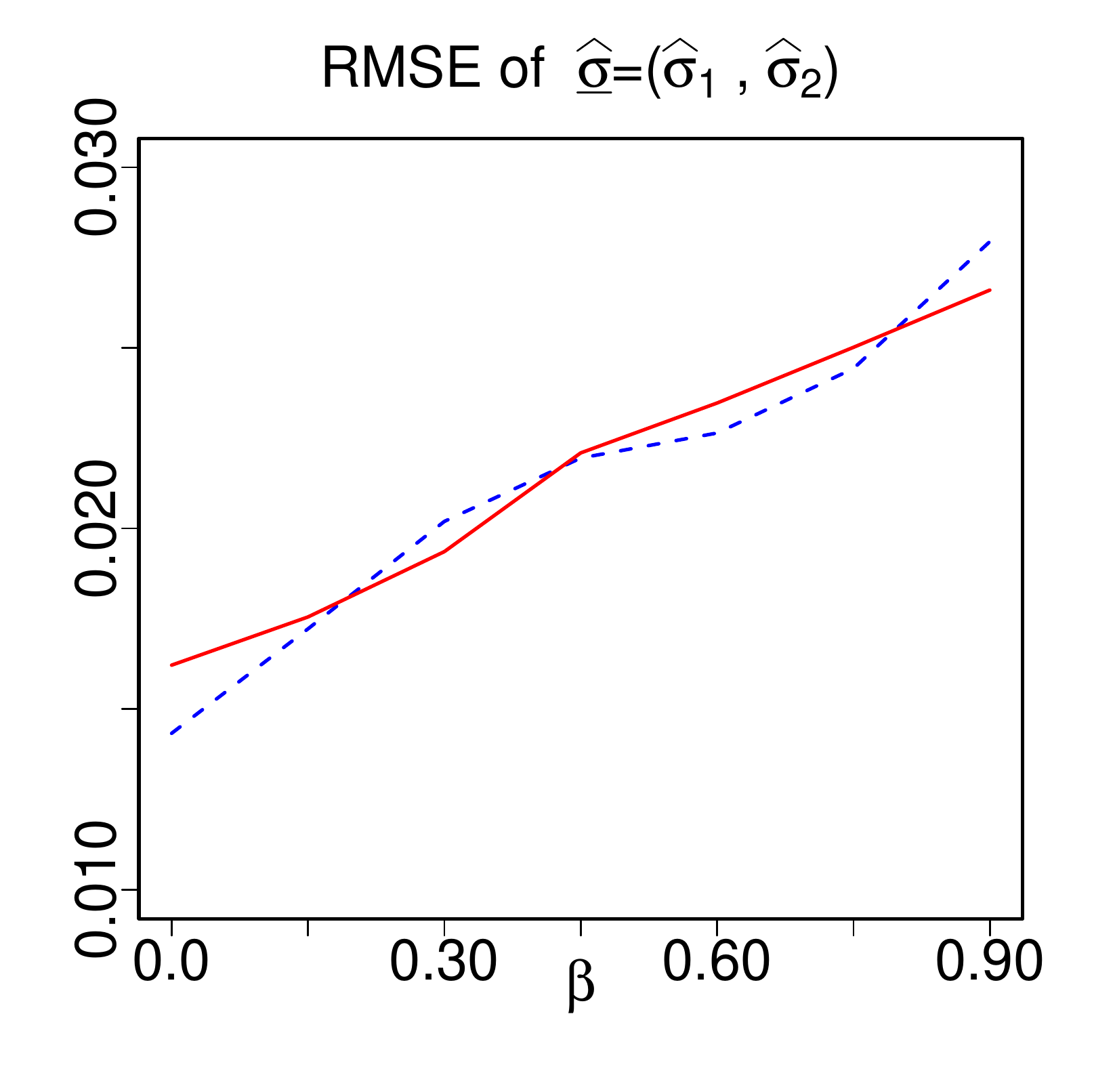}&
\includegraphics[width=40mm,height=25mm]{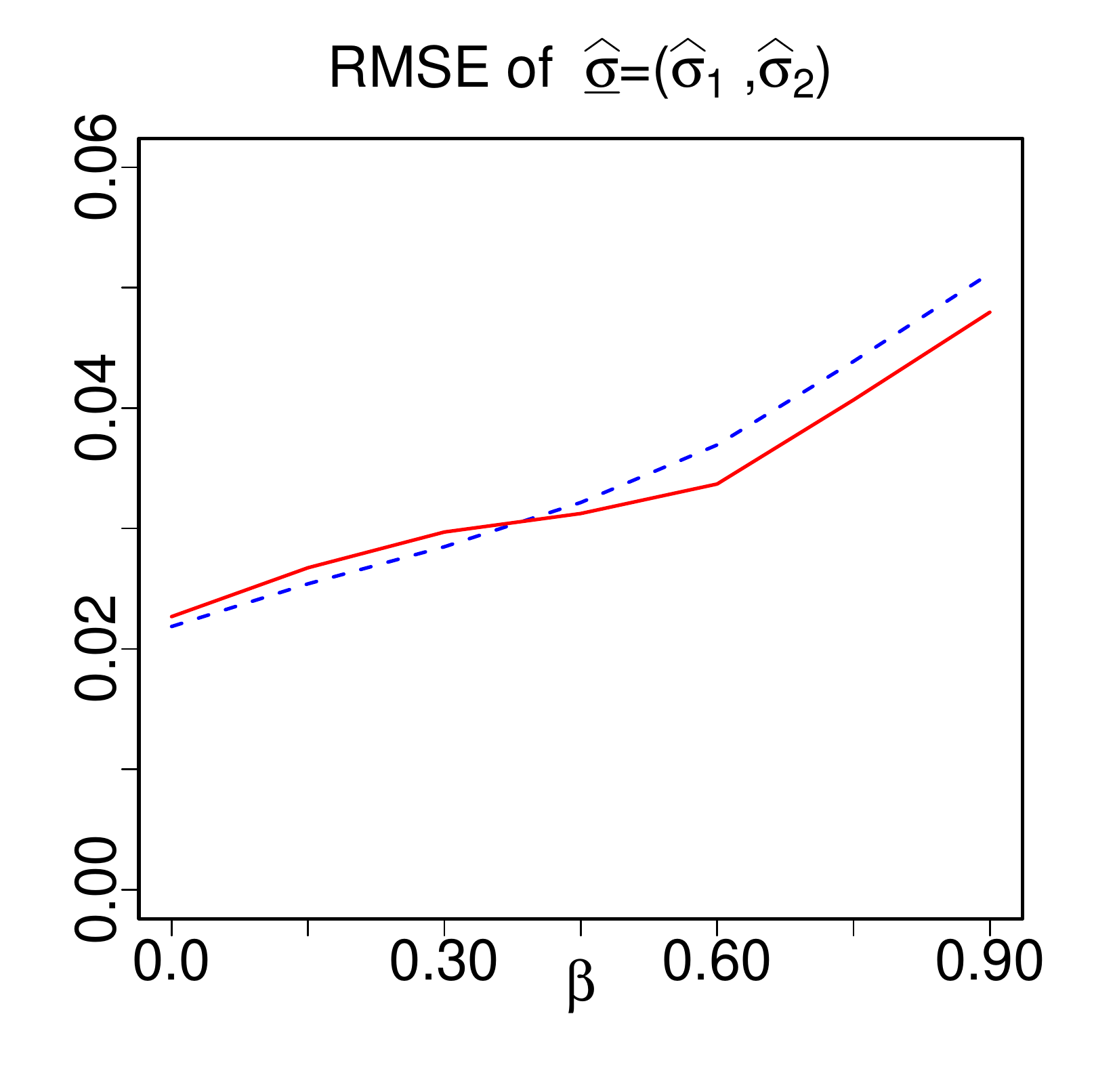}\\
\includegraphics[width=40mm,height=25mm]{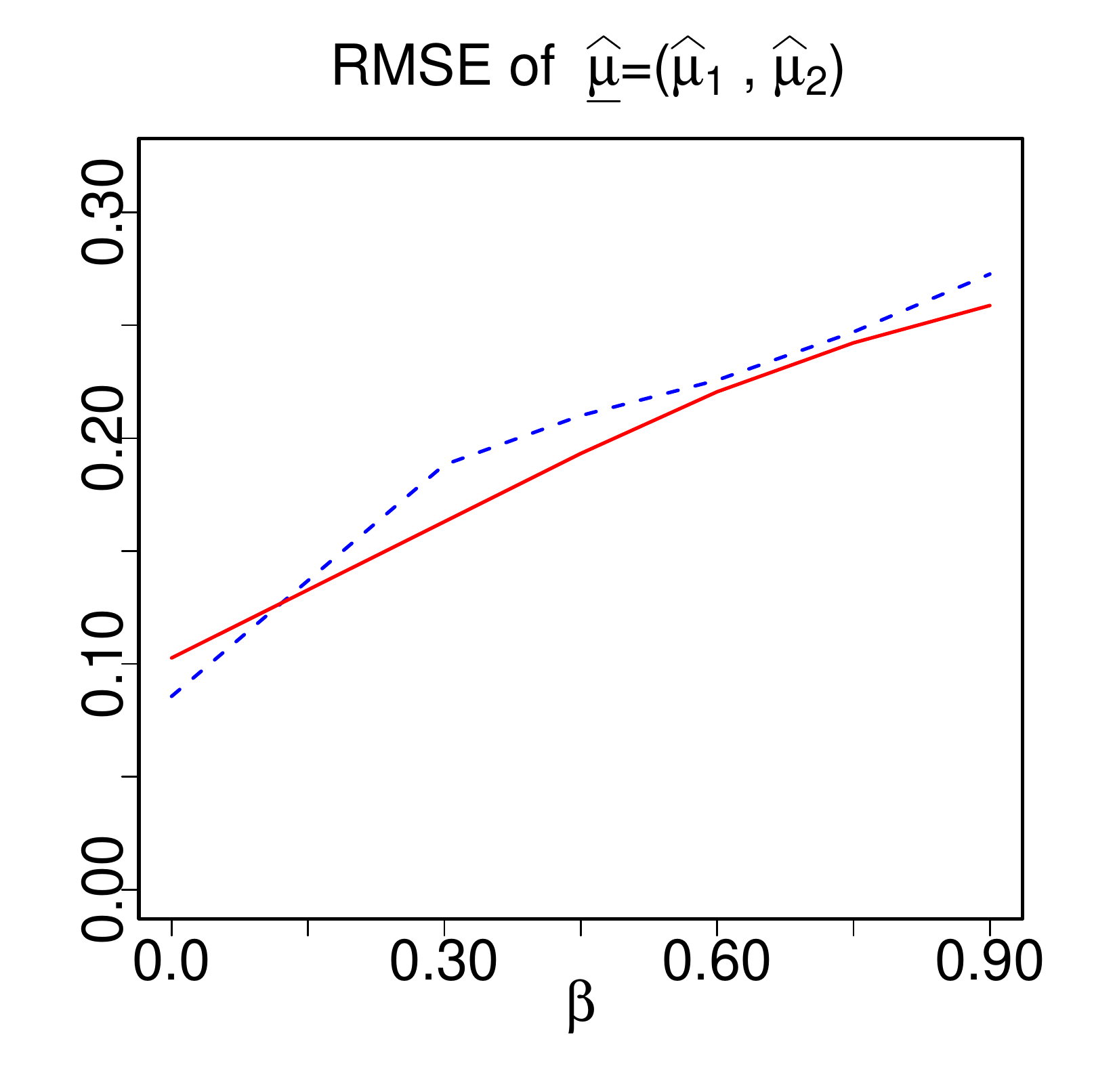}&
\includegraphics[width=40mm,height=25mm]{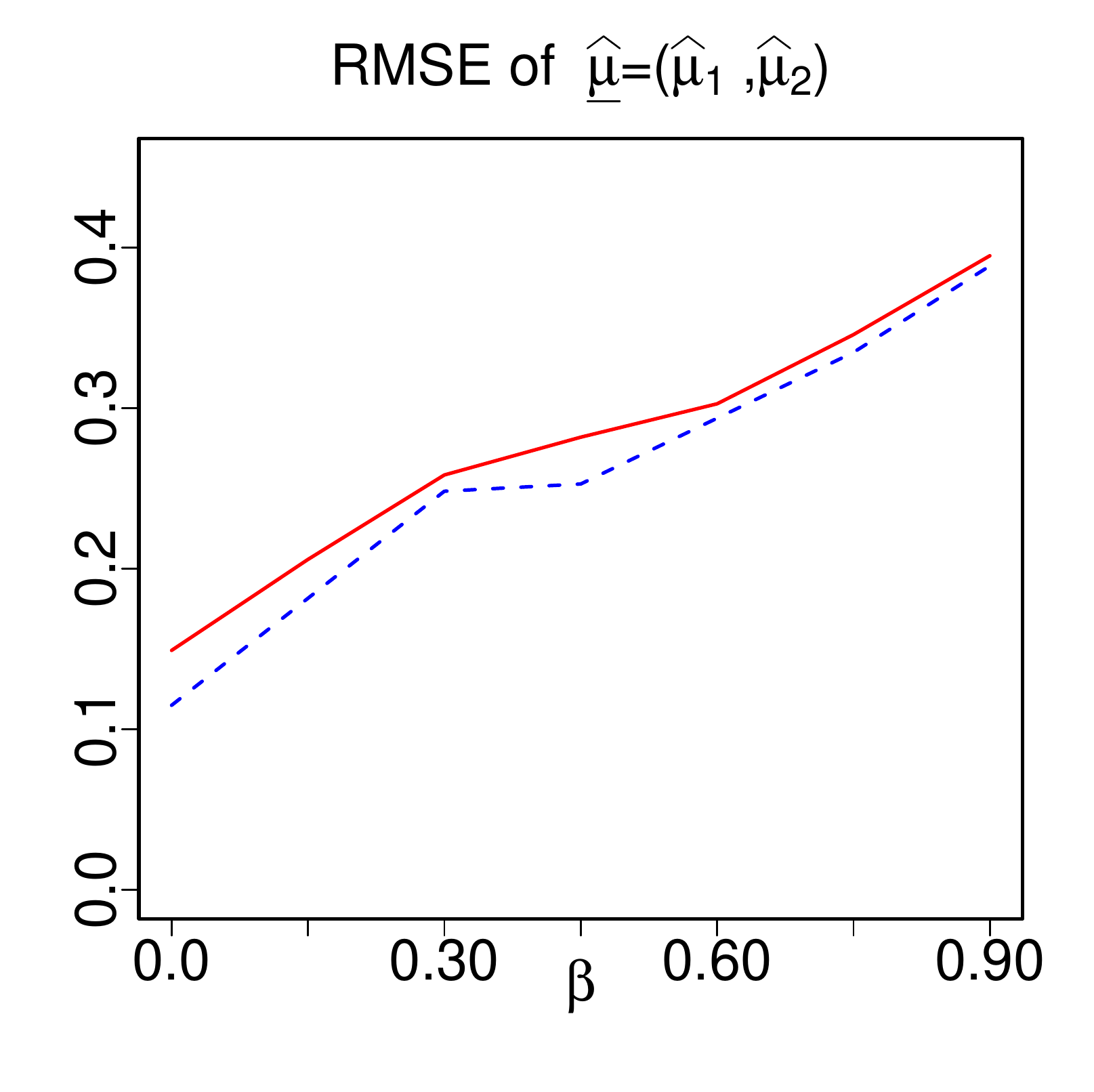}\\
\end{tabular}}
\caption{The RMSE of $\underline{\hat{\omega}}$, $\underline{\hat{\beta}}$, $\underline{\hat{\sigma}}$, and $\underline{\hat{\mu}}$ when the EM algorithm is applied to the sample of size 1000 generated from ${\cal{M}}C_{0}\bigl(\underline{\mu},\underline{\eta},\underline{\lambda},2\bigr)$ for 200 replications. In each sub-figure, dashed red line and blue solid line refer to RMSE of the estimator of the first and the second components, respectively. The levels of the skewness parameter on the horizontal axis are 0.0,0.15,0.30,0.45,0.60,0.75, and 0.90. The left-hand side and right-hand side sub-figures correspond to the scale vector $\underline{\sigma}=(0.25,0.25)$ and $\underline{\sigma}=(0.5,0.5)$, respectively.}
\label{fig2}
\end{figure}

Here, we perform a simulation study to investigate the performance of the EM algorithm in estimating the parameters of two-component Cauchy mixture model.
For this, we generate a sample of size 1000 for 200 times under two scenarios given by the following.
\begin{enumerate}
\item $\underline{\beta}=(\beta_1,\beta_2)=(\beta,\beta)$ for $\beta=0.0,0.15,0.30,0.45,0.60,0.75,0.90$, $\underline{\sigma}=(\sigma_1,\sigma_2)=(0.25,0.25)$,
$\underline{\mu}=(\mu_1,\mu_2)=(-3,3)$, and $\underline{\omega}=(\omega_1,\omega_2)=(0.5,
0.5)$. 
\item $\underline{\beta}=(\beta_1,\beta_2)=(\beta,\beta)$ for $\beta=0.0,0.15,0.30,0.45,0.60,0.75,0.90$, $\underline{\sigma}=(\sigma_1,\sigma_2)=(0.5,0.5)$, $\underline{\mu}=(\mu_1,\mu_2)=(-3,3)$, and $\underline{\omega}=(\omega_1,\omega_2)=(0.5,0.5)$. 
\end{enumerate}
The results of simulations are displayed in Figure \ref{fig2}. 

\subsection{Model validation using the real data}
\begin{figure}
\resizebox{\textwidth}{!}
{\begin{tabular}{cc}
\includegraphics[width=40mm,height=40mm]{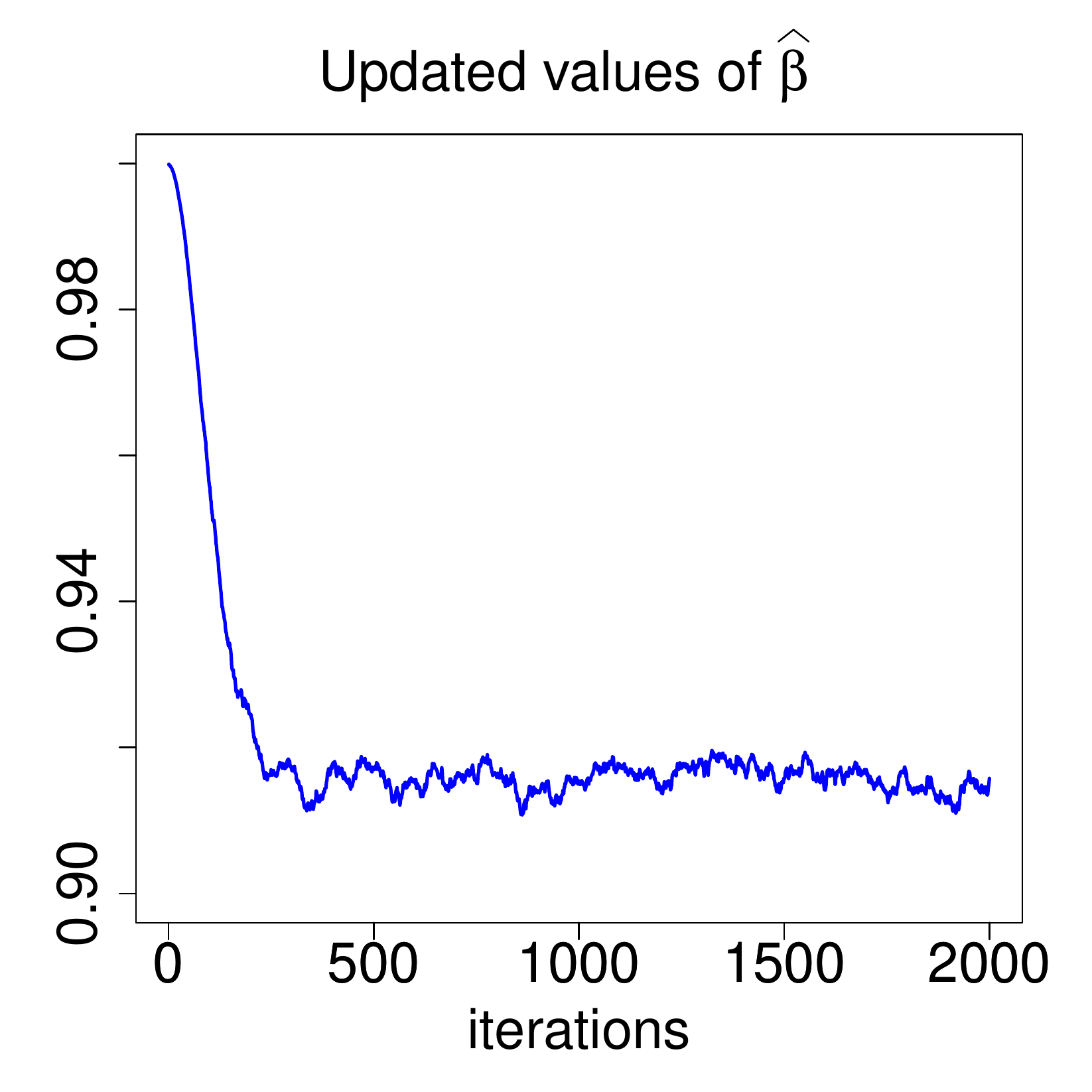}&
\includegraphics[width=40mm,height=40mm]{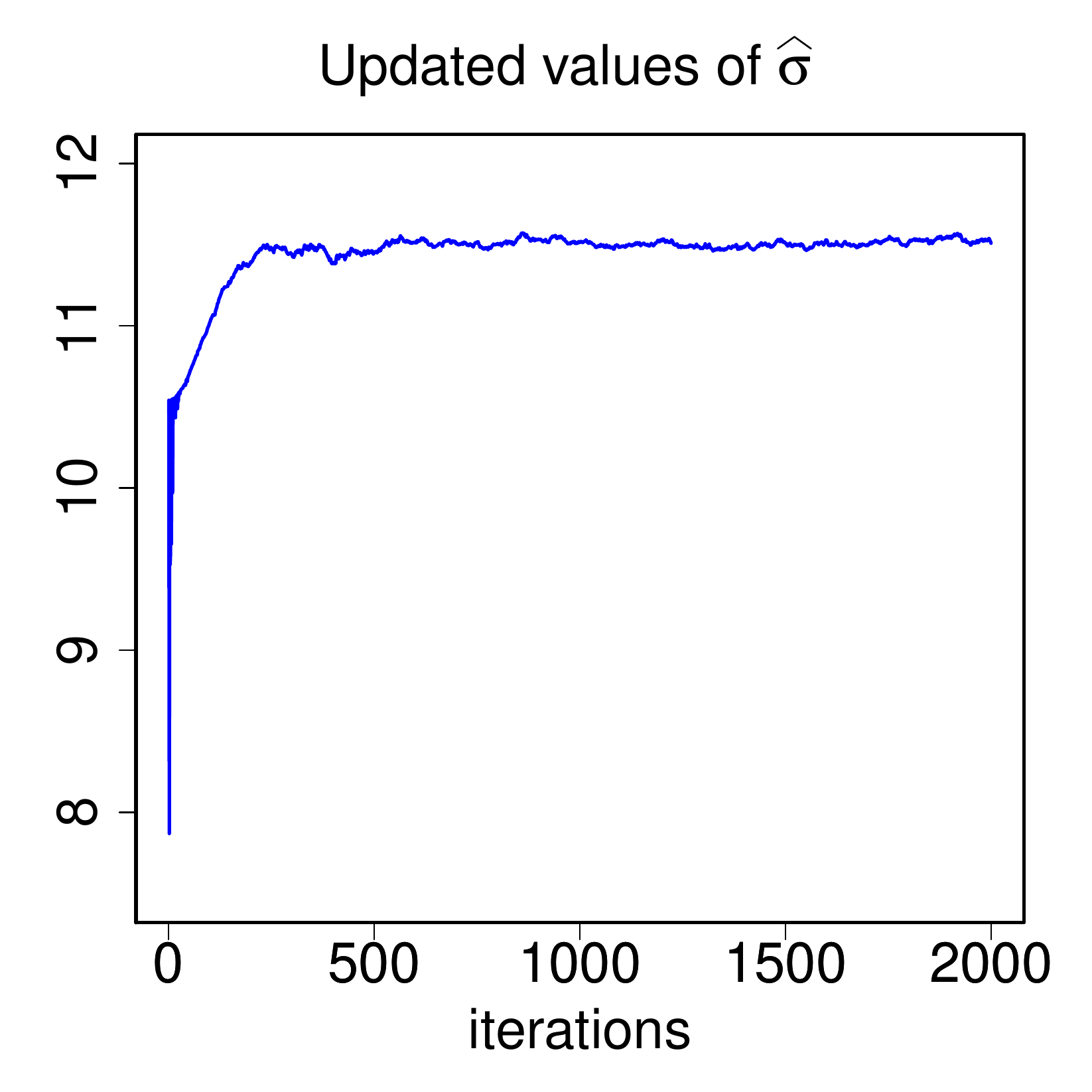}\\
\includegraphics[width=40mm,height=40mm]{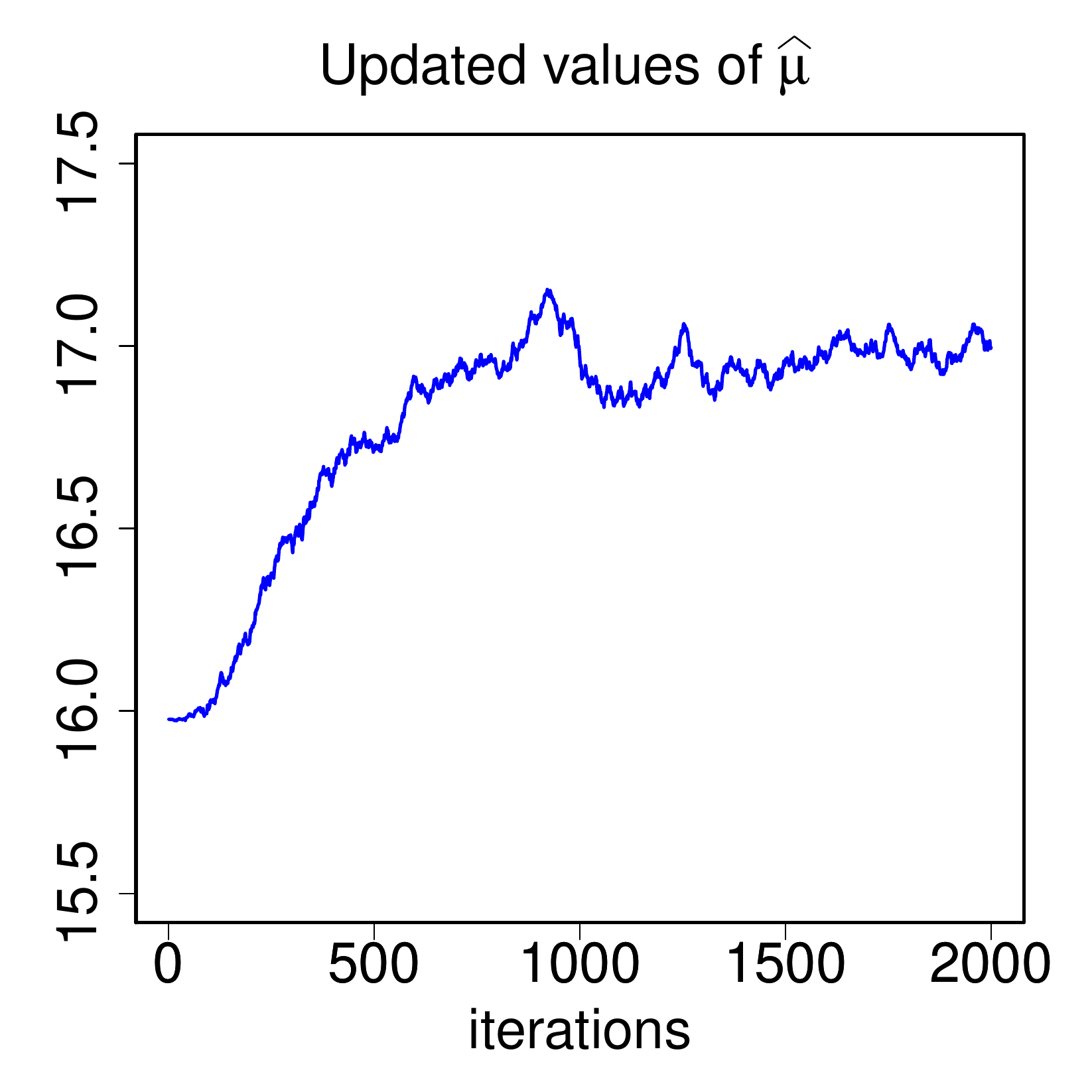}&
\end{tabular}}
\caption{The time series graphs of the updated parameters $\hat{\beta}$, $\hat{\sigma}$, and $\hat{\mu}$ when the EM algorithm is applied to the large recorded intensities (in Richter scale) of the earthquake data.}
\label{fig3}
\end{figure}

Here, we consider the large recorded intensities (in Richter scale) of the earthquake in seismometer locations in western North America between 1940 and 1980. The related features was reported by \cite{Joyner1981} and also analyzed by \cite{Davidian1995}. Among the features, we focus on the 182 distances from the seismological measuring station to the epicenter of the earthquake (in km) as the variable of interest. To implement the EM algorithm, we used the quantile-based estimation as the initial values, see \cite{McCulloch1986}. The time series graphs of the updated parameters versus 2000 iterations are displayed in Figure \ref{fig3}. As it is seen, the convergence occurs after a large numbers of iterations. By averaging the updated values between 1000-th and 2000-th iterations, the EM-based estimations for the parameters are: $\hat{\mu}=16.9498$, $\hat{\sigma}=11.5981$, and $\hat{\beta}=0.9167$. The Kolmogorov-Smirnov (K-S) goodness-of-fit and Anderson-Darling (A-D) criteria are 0.0397 and 0.6980, respectively.

\begin{figure}[h!]
\resizebox{\textwidth}{!}
{\begin{tabular}{cc}
\includegraphics[width=40mm,height=40mm]{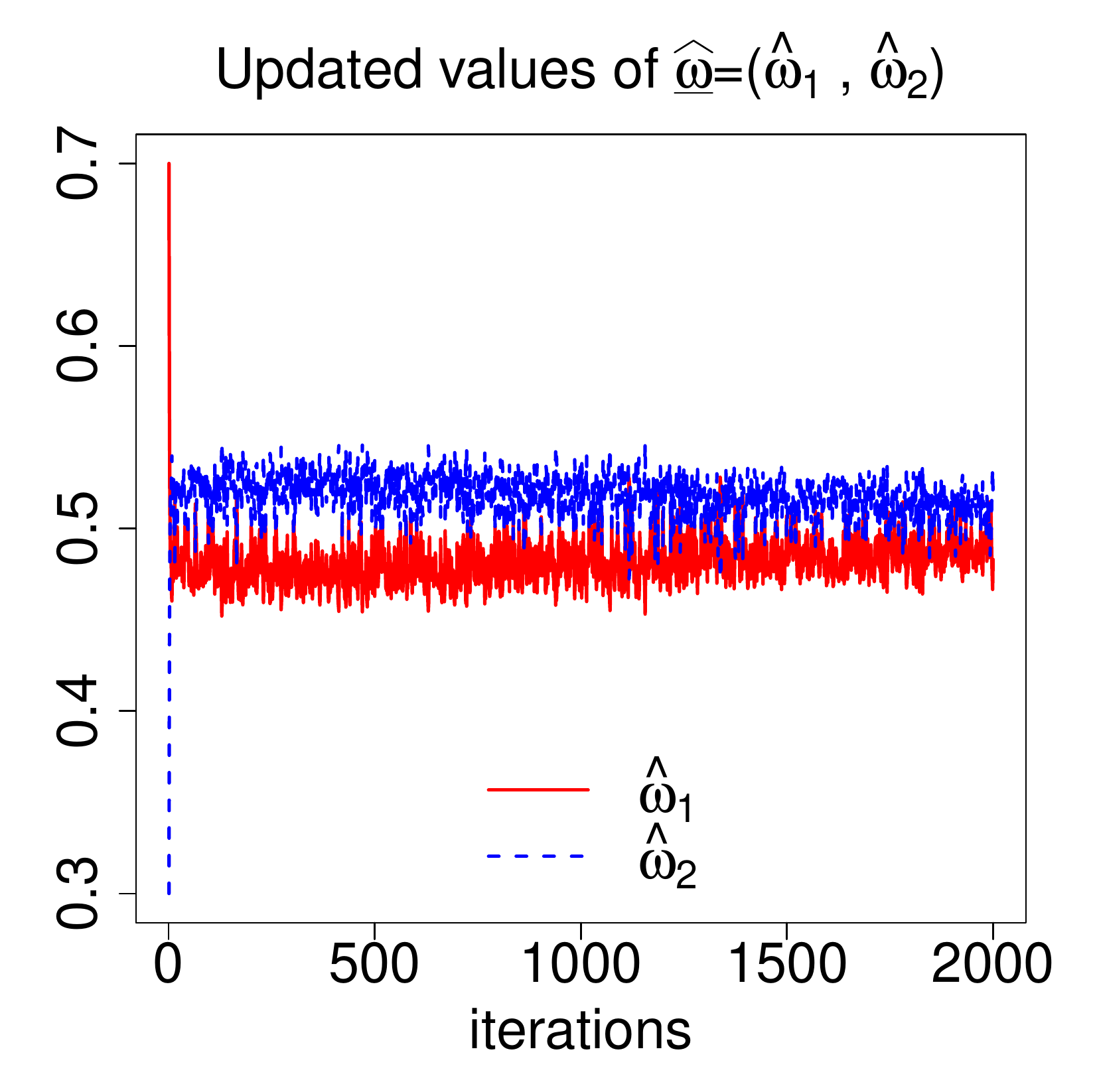}&
\includegraphics[width=40mm,height=40mm]{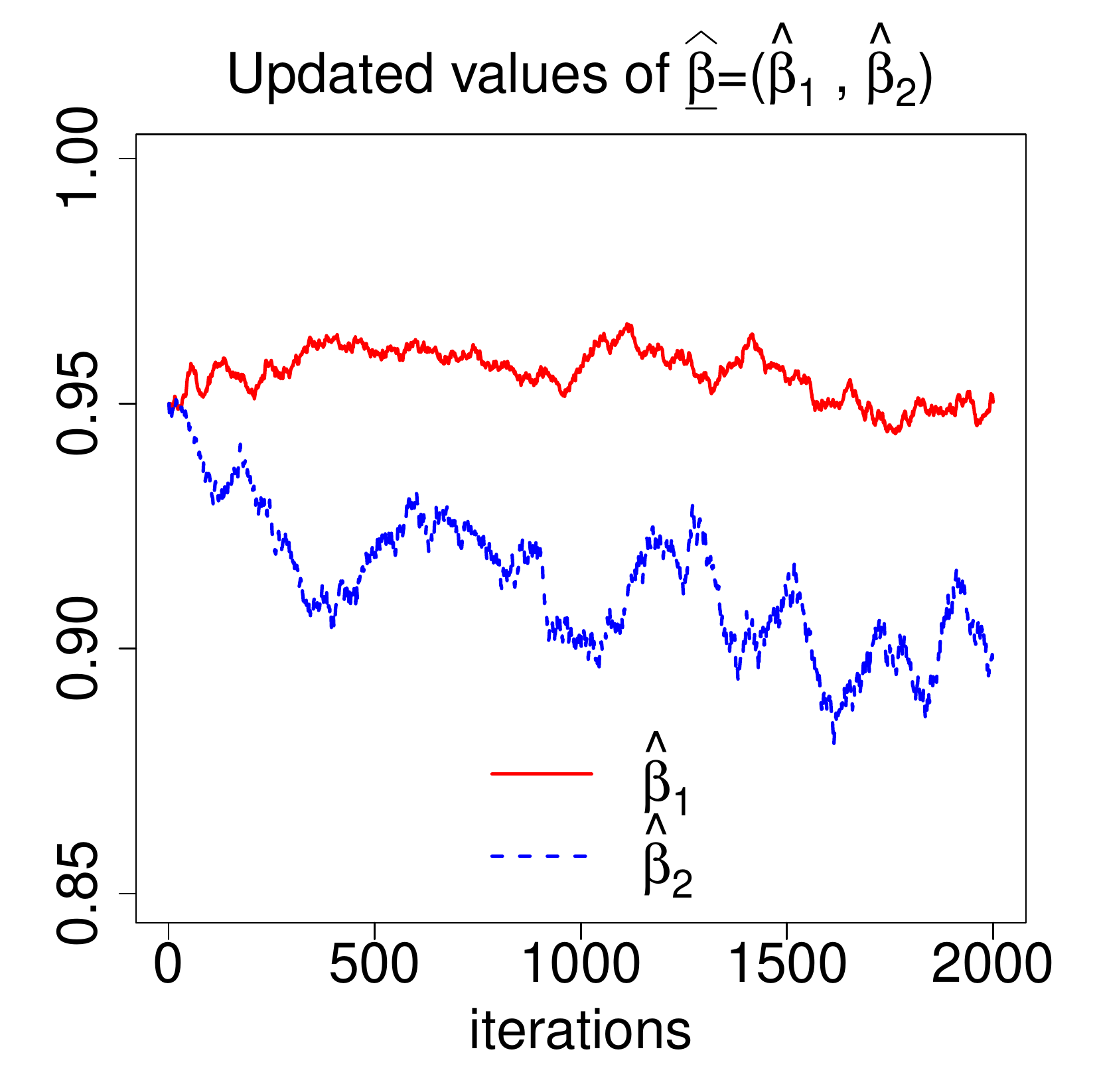}\\
\includegraphics[width=40mm,height=40mm]{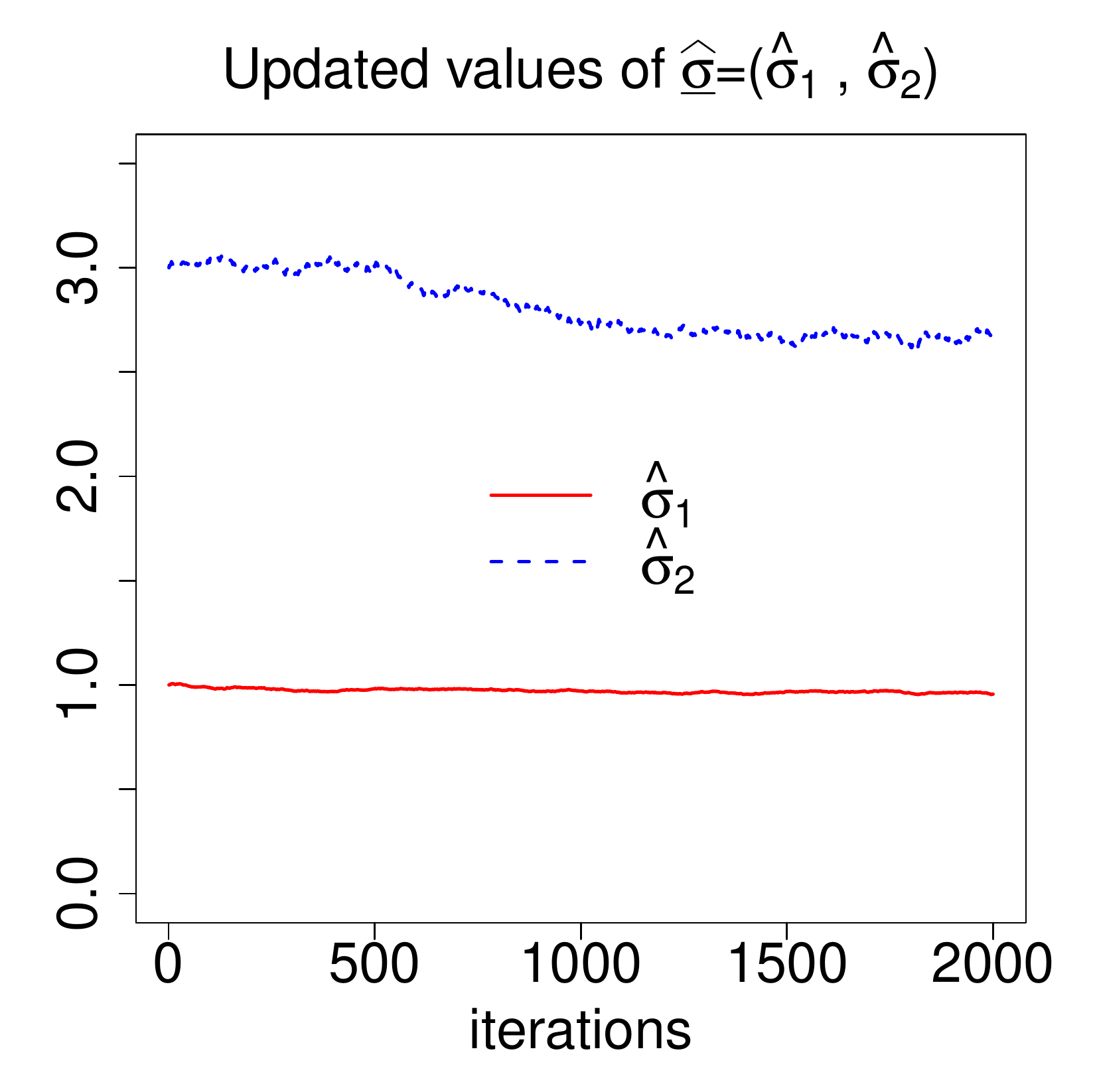}&
\includegraphics[width=40mm,height=40mm]{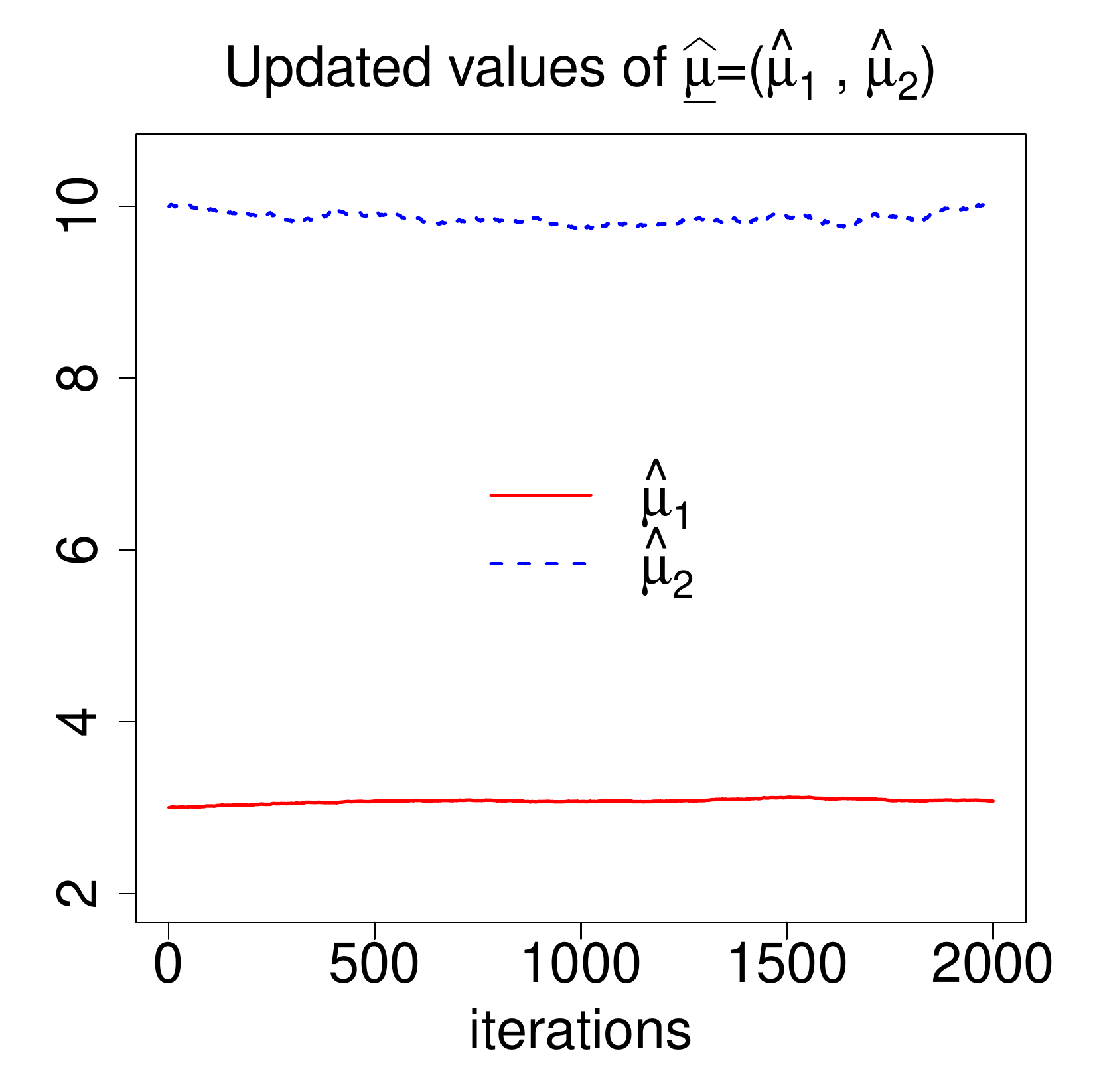}\\
\end{tabular}}
\caption{Time series graphs of the updated parameters $\hat{\underline{\omega}}=(\hat{\omega}_1, \hat{\omega}_2)$, $\hat{\underline{\beta}}=(\hat{\beta}_1, \hat{\beta}_2)$, $\hat{\underline{\sigma}}=(\hat{\sigma}_1, \hat{\sigma}_2)$, and $\hat{\underline{\mu}}=(\hat{\mu}_1, \hat{\mu}_2)$ when the EM algorithm is applied to the tetrahydrocortisone data.}
\label{fig4}
\end{figure}
\begin{figure}[h!]
\resizebox{\textwidth}{!}
{\begin{tabular}{cc}
\includegraphics[width=40mm,height=40mm]{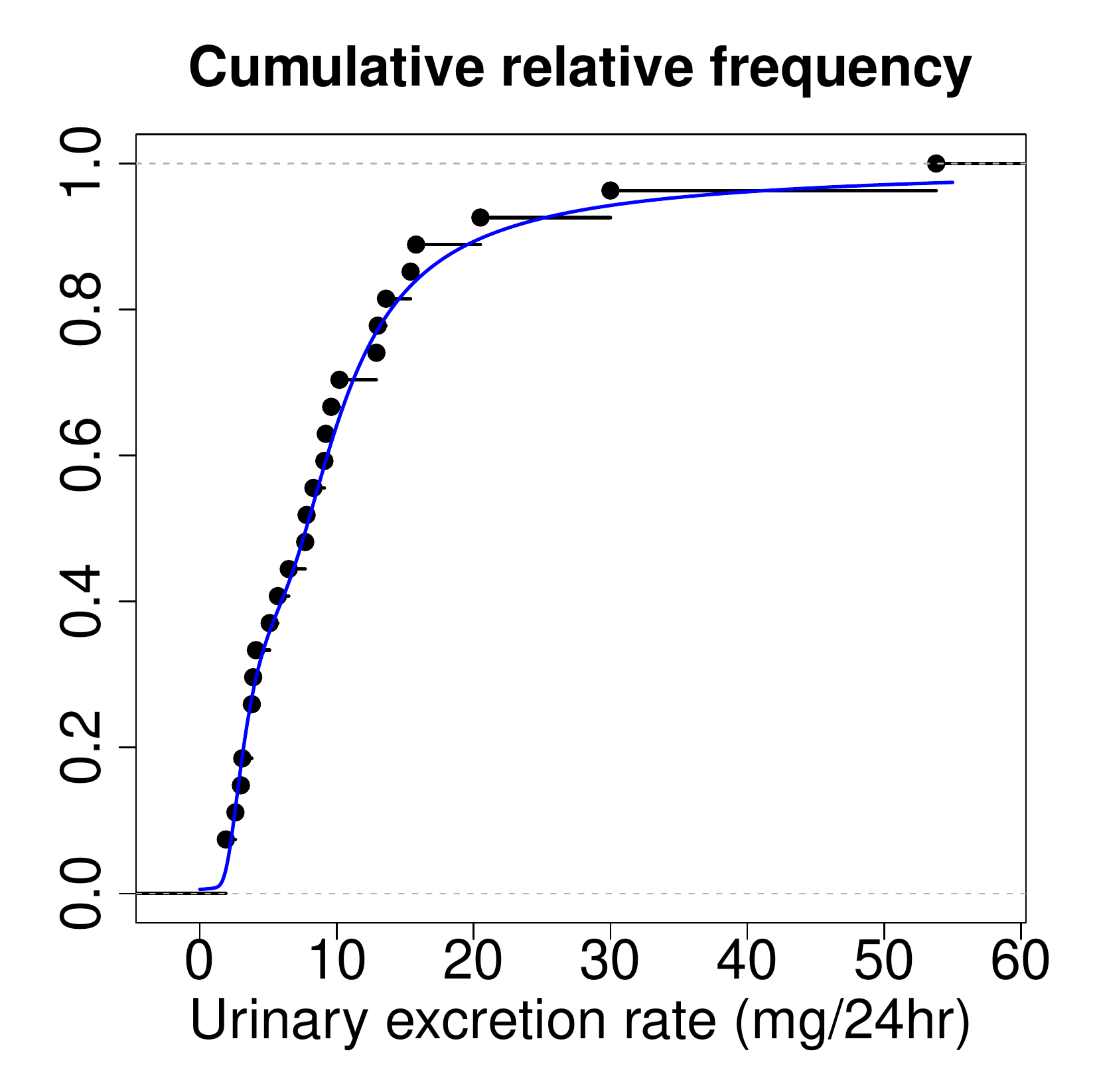}&
\includegraphics[width=40mm,height=40mm]{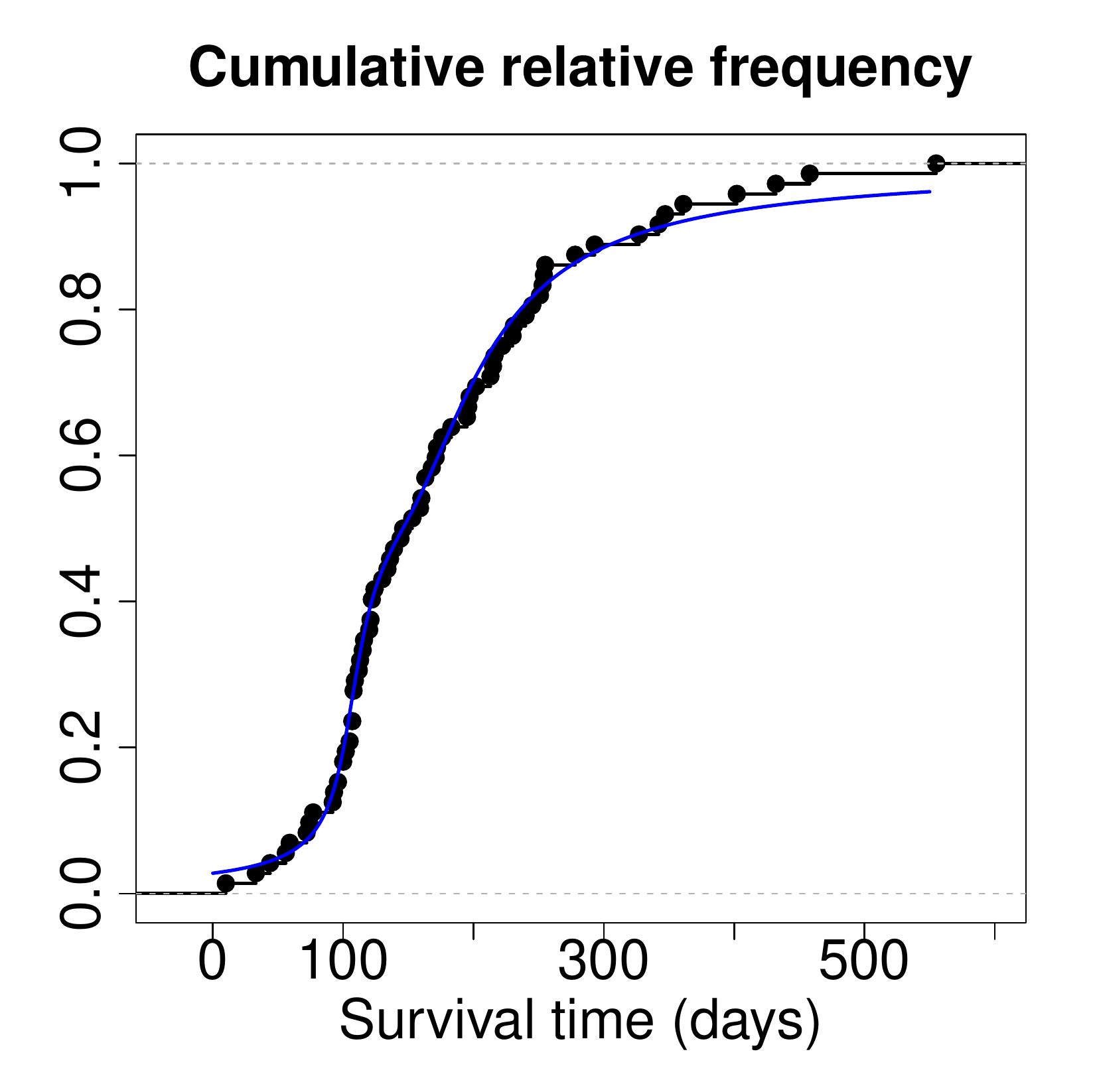}\\
\end{tabular}}
\caption{Empirical distribution functions of urinary excretion rate of tetrahydrocortisone (left-hand side) and survival time (right-hand side) data. In each sub-figure, the fitted cdf of a two-component mixture of Cauchy distributions is shown by a blue solid curve.}
\label{fig5}
\end{figure}
\subsection{Mixture Model validation using the real data}
Application of the mixture of Cauchy distributions will be illustrated by two sets of data. The first set is related to the diagnostic tests on patients with cushing's syndrome, see \cite{Venables2002}, and the second set is the survival time in days of 72 guinea pigs infected with virulent tubercle bacilli, see \cite{Bjerkedal1960}. For cushing's syndrome data, we focus on the urinary excretion rate (mg/24hr) of tetrahydrocortisone. We apply the EM algorithm described in Section 3 to estimate the parameters of two-component Cauchy mixture model applied to this set of data. For implementing the EM algorithm, the initial values are chosen as: $\underline{\omega}_{0}=(0.70,0.30)$, $\underline{\sigma}_{0}=(1,3)$, $\underline{\beta}_{0}=(0.95,0.95)$, and $\underline{\mu}_{0}=(3,10)$. 
The time series graphs of the updated parameters based on 2000 iterations are displayed in Figure \ref{fig4}. As it is seen, the initial values are started well away from the EM-based estimations. The fitted cdf to the urinary excretion rate data are shown in left hand-side of subfigure of Figure \ref{fig5}. For this set of data, the weight, scale, skewness, and the location vectors are estimated as $\underline{\hat{\omega}}=(0.485,0.515)$, $\underline{\hat{\sigma}}=(0.964,2.681)$, $\underline{\hat{\beta}}=(0.954,0.904)$, and $\underline{\hat{\mu}}=(3.089,9.852)$, by averaging the updated values between 1000-th and 2000-th iterations. As it is seen from Figure \ref{fig5}, the fitted cdf captures well the general shape of the empirical distribution function. The corresponding K-S and A-D statistics are 0.083 and 0.147, respectively. 
\par For survival time of guinea pigs, we used initial values $\underline{\omega}_{0}=(0.65,0.35)$, $\underline{\sigma}_{0}=(20,55)$, $\underline{\beta}_{0}=(0.20,0.05)$, and $\underline{\mu}_{0}=(110,250)$. By averaging the updated values between 1000-th and 2000-th iterations, the EM-based estimators are $\underline{\hat{\omega}}=(0.547,0.453)$, $\underline{\hat{\sigma}}=(16.397,39.112)$, $\underline{\hat{\beta}}=(0.112,0.803)$, and $\underline{\hat{\mu}}=(107.865,196.269)$. The corresponding K-S and A-D statistics are 0.0599 and 0.301, respectively. The fitted cdf to the survival time of guinea pigs are shown in right-side of subfigure of Figure \ref{fig5}.
\section{Conclusion}
We have derived the estimators of the parameters of Cauchy and mixture of Cauchy distributions using the EM algorithm. In the both cases, performance of the EM algorithm have been demonstrated in the sense of root of mean square error through simulations. The performance of the EM algorithm has been shown by applying it to the real data. Since the Monte Carlo approximations have been used to evaluate the required expectation in M-step of the EM algorithm, time series plots of the updated parameters do not converge to the global maximum point, but instead, it converges to the true distribution. Therefore, as a rule of thumb, we obtain the EM-based estimators by averaging the updated values between 500-th and 1000-th iterations. Our analyses reveal that the proposed EM algorithm is robust with respect to initial values and they can be chosen well away from their true values.
\par For introducing a multivariate Cauchy distribution, there are a variety of possible generalizations. As the first possible future work, the EM algorithm can be applied to a $d$-dimensional Cauchy random vector ${\bf{Y}}$ which is represented as
\begin{align}
{\bf{Y}} =\frac{{\bf{N}}}{{\bf{Z}}}+ {\bf{\Lambda}} {\bf{V}}+{\boldsymbol{\delta}},\nonumber
\end{align}
where ${{\bf{N}}}$ is a multivariate normal random vector with zero location vector and covariance matrix $\Sigma$ whose main diagonal entries are of the form $\Sigma_{ii}=\sigma_{i}(1-|\beta_{i}|)$, ${\bf{\Lambda}}=(\sigma_{1}\beta_{1},\dots,\sigma_{d} \beta_{d})^{T}$, ${{\bf{Z}}}=(Z_1,\dots,Z_d)^{T}$ independent of ${{\bf{N}}}$, follows a multivariate normal distribution with zero location vector and unity diagonal covariance matrix, ${{\bf{V}}}=(V_1,\dots,V_d)^{T}$ independent of ${{\bf{N}}}$ and ${{\bf{Z}}}$ is a vector of $d$ independent random variables that each follows $C_{0}(1,1,0)$, and ${\boldsymbol{\delta}}=(\mu_{1}+\sigma_{1}\beta_{1} \log |\sigma_{1}\beta_{1}|,\dots,\mu_{d}+\sigma_{d}\beta_{d} \log |\sigma_{d}\beta_{d}|)^{T}$ is the location vector. Here, $\sigma_{i}$ and $\beta_{i}$; for $i=1,\dots,d$, are the marginal scale and skewness parameters, respectively. Besides, by virtue of representation (\ref{prop11}), a Bayesian approach can be developed to estimate the parameters of $C_{0}(\beta,\sigma,\mu)$ distribution. Programs in $\mathsf{R}$ language for implementing the EM algorithm can be obtained from the author upon request.
\\\\\\
{\Large{\bf{Appendix~ A: Proof ~of~ Proposition~ 1}}}\\\\
Let $T$ and $P$ denote two independent random variables which follow $C_{1}(0,1,0)$ and $C_{0}(1,1,0)$. Define $Y=\eta T+\lambda P+\delta$ in which $\eta=\sigma\left(1-|\beta|\right)$, $\lambda=\sigma \beta$, and $\delta=\mu+\frac{2}{\pi}\lambda \log |\lambda|$. We can write
\begin{align}\label{lab1}
E \exp (jtY)=&E \exp \biggl\{jt\bigl(\sigma\left(1-|\beta|\right) T+\lambda P+\mu +\frac{2}{\pi}\sigma \beta \log |\sigma \beta|\bigr)\biggr\} \nonumber \\
=&E \exp \Bigl\{jt\sigma\left(1-|\beta|\right) T\Bigl\} \times E \exp \Bigl\{jt\Bigl(\sigma \beta P+\mu+\frac{2}{\pi}\sigma \beta \log |\sigma \beta|\Bigr)\Bigl\}\nonumber\\
=&\exp\Bigl\{-(1-|\beta|)\left| \sigma t \right|\Bigl\} \times E \exp \Bigl\{jt\Bigl(\sigma \beta P+\mu+\frac{2}{\pi}\sigma \beta \log |\sigma \beta|\Bigr)\Bigl\}\nonumber\\
=&\exp\Bigl\{-(1-|\beta|)\left| \sigma t \right|\Bigl\} \times {\cal{E}}.
\end{align}
Since $P$ comes from $C_{0}(1,1,0)$, then $\sigma \beta P+\mu+\frac{2}{\pi}\sigma \beta \log |\sigma \beta|$ follows $C_{0}\bigl(\mathrm{sgn}(\beta),|\sigma \beta|,\mu+\frac{2}{\pi} \sigma \beta \log |\sigma \beta|\bigr)$, see \cite[pp. 190]{Nolan1998}. It follows from (\ref{chf}) that
\begin{align}\label{lab2}
{\cal{E}}=\exp\biggl\{-\left| \sigma \beta t \right|\biggl[1+j\frac{2}{\pi}\mathrm{sgn}(t \beta)\log \left| t \sigma \beta\right|\biggr]+j t \mu+j t\frac{2}{\pi} \sigma \beta \log |\sigma \beta|\biggr\}.
\end{align}
Substituting $\lambda=\beta \sigma $ the right hand-side of (\ref{lab2}) and simplifying, we have
\begin{align}\label{lab3}
{\cal{E}}=&\exp\biggl\{-\left| \sigma \beta t \right|-j\frac{2}{\pi}\beta\mathrm{sgn}(t) |\sigma t| \log |\sigma \beta t|+j t \mu+j t\frac{2}{\pi} \beta \sigma \log |\beta \sigma|\biggr\}\nonumber\\
=&\exp\biggl\{-\left| \sigma \beta t \right|-j\frac{2}{\pi}\beta\mathrm{sgn}(t) |\sigma t| \log |t|-j\frac{2}{\pi} \beta\mathrm{sgn}(t) |\sigma t| \log |\beta \sigma|+j t \mu+j t\frac{2}{\pi} \beta \sigma \log |\beta \sigma|\biggr\}\nonumber\\
=&\exp\biggl\{-\left| \sigma \beta t \right|-j\frac{2}{\pi}\beta\mathrm{sgn}(t) |\sigma t| \log |t|+j t \mu \biggr\}.
\end{align}
By replacing right hand-side of (\ref{lab3}) with ${\cal{E}}$ in (\ref{lab1}), we get
\begin{align}
E \exp (jtY)=\exp\biggl\{-| \sigma t |\Bigl[1+j\frac{2}{\pi}\beta\mathrm{sgn}(t) \log |t|\Bigr]+j t \mu \biggr\},\nonumber 
\end{align}
where the last expression is the chf of $Y$ which follows $C_{1}(\beta,\sigma,\mu)$. Since $T$ is a standard Cauchy random variable, it is well known that the random variable $T$ can be represented as the ratio of two independent standard normal random variables, $N$ and $Z$ say. Therefore
$Y=\eta \frac{N}{Z}+\lambda P+\delta$ follows $C_{1}(\beta,\sigma,\mu)$. The proof is complete. \\\\\\
{\Large{\bf{Appendix~ B: Proof ~of~ Proposition~ 2}}}\\\\
Without loss of generality, suppose $I=E\bigl(Z^{2}P^{r}\big|y,\Theta\bigr)$; for $r=0,1,2$. It follows from (\ref{rep2}), that the joint pdf of $Z$ and $P$ given $y$ and $\Theta$, i.e., $q(z, p|y,\Theta)$ takes the form
\begin{align}\label{I1}
q(z, p|y,\Theta)=\frac{g\big(y\big|\lambda p+\delta,\eta^2/z^2\big)g(z|0,1)h(p)}{f(y|\Theta)}
\end{align}
where the pdfs $g$, $h$, and $f$ in the right-hand side of (\ref{I1}) are defined after representation (\ref{rep1}). We can write
\begin{align}\label{I2}
I=\frac{1}{f\bigl(y\big|\Theta\bigl)}\int_{\rm I\!{R}} \int_{\rm I\!{R}}z^{2}p^{r}g\big(y\big|\lambda p+\delta,\eta^2/z^2\big)g(z|0,1)h(p)dzdp.
\end{align}
After substituting the algebraic form of the corresponding pdfs in the right-hand side of (\ref{I2}) and simplifying, we have 
\begin{align}
I=\frac{1}{f\bigl(y\big|\Theta\bigl)}\int_{\rm I\!{R}}\frac{p^{r}h(p)}{\pi \eta} \int_{\rm I\!{R}^{+}}z^{3} e^{-\frac{z^2(1+q^2)}{2}}dzdp,\nonumber
\end{align}
where $q=\bigl(y-\lambda p-\delta\bigr)/\eta$. Making a change of variable of the form $z^2(1+q^2)/2=w$ yields
\begin{align}
I&=\frac{2}{f\bigl(y\big|\Theta\bigl)}\int_{\rm I\!{R}}\frac{p^{r}h(p)}{\pi \eta(1+q^2)^{2}} \int_{\rm I\!{R}^{+}}w e^{-w}dwdp\nonumber\\
&=\frac{2}{\pi \eta f\bigl(y\big|\Theta\bigl)}\int_{\rm I\!{R}}\biggl[1+\biggl(\frac{y-\lambda p-\delta}{\eta}\biggr)^{2}\biggr]^{-2}p^{r}h(p)dp,\nonumber
\end{align}
where $r=0,1,2$. Since $h(p)$ has no closed-form expression, we approximate $I$ through the Monte Carlo method. For this, we generate 3000 realizations from random variable $P$ which follows $C_{0}(1,1,0)$ using $\mathsf{STABLE}$ software based on method developed by
\cite{Chambers1976}. So, a fair approximation of $I$ is given by
\begin{align}\label{approx1}
I&\approx \frac{1}{1500\pi \eta f\bigl(y\big|\Theta\bigl)}\sum_{j=1}^{3000}\biggl[1+\biggl(\frac{y-\lambda p_{j}-\delta}{\eta}\biggr)^{2}\biggr]^{-2}p_{j}^{r}.\nonumber
\end{align}
We note that $f\bigl(y\big|\Theta\bigl)$ in denominator of approximation is computed using $\mathsf{STABLE}$ software.


\begin{thebibliography}{00}
\bibitem{Bjerkedal1960} 
Bjerkedal, T. (1960). Acquisition of resistance in guinea pigs infected with different doses of virulent tubercle bacilli, \emph{American Journal of Epidemiology}, 72, 130-148.
\bibitem{Casarin2004} 
Casarin, R. (2004). Bayesian inference for mixtures of stable distributions, working paper No. 0428, CEREMADE, University Paris IX. 
\bibitem{Chambers1976} 
Chambers, J., Mallows, C., and Stuck, B. (1976). A method for simulating stable random variables, \emph{Journal of the American Statistical Association}, 71, 340-344.
\bibitem{Davidian1995}
Davidian, M. and Giltinan, D. M. (1995). \emph{Nonlinear Models for Repeated Measurement Data}, Chapman and Hall, London. 
\bibitem{Dempster1977} 
Dempster, A. P., Laird, N. M., and Rubin, D. B. (1977). Maximum likelihood from incomplete data via the EM algorithm, \emph{Journal of the Royal Statistical Society Series B}, 39, 1-38. 
\bibitem{Johnson2004} 
Johnson, O. (2004). \emph{Information theory and the central limit theorem}, Imperial College Press, London.
\bibitem{Joyner1981} 
Joyner, W. B. and Boore, D. M. (1981). Peak horizontal acceleration and velocity from strong-motion records including records from the 1979 Imperial Valley, California, earthquake, \emph{Bulletin of the Seismological Society of America}, 71, 2011-2038.
\bibitem{Klebanov2006} 
Klebanov, L. B. and Kozubowski, T. J., and Rachev, S. T. (2006). \emph{Ill-posed problems in probability and stability of random sums}, Nova Science Publishers, New York.
\bibitem{Kagan1992} 
Kagan, Y. Y. (1992). Correlations of earthquake focal mechanisms, \emph{Geophysical Journal International}, 110, 305-320.
\bibitem{Lorentz1906} 
Lorentz, H. A. (1906). The absorption and emission lines of gaseous bodies, \emph{Proceedings of the Royal Netherlands Academy of Arts and Sciences}, 591-611.
\bibitem{Maechler2015} 
Maechler, M., Rousseeuw, P., Struyf, A., Hubert, M., and Hornik, K. (2015). {\it{\textsf{cluster}}}: Cluster Analysis Basics and Extensions, {\it{\textsf{R}}} package version 2.0.1. 
\bibitem{McLachlan2008} 
McLachlan, G. J. and Krishnan, T. (2008). \emph{The EM Algorithm and Extensions}, second edition, John Wiley. 
\bibitem{McCulloch1986} 
McCulloch, J. H. (1986). Simple consistent estimators of stable distribution parameters, \emph{Communications in Statistics-Simulation and Computation}, 5, 1109-1136.
\bibitem{Min1996} 
Min, I. A., Mezic, I. and Leonard, A. (1996). Levy stable distributions for velocity and velocity difference in systems of vortex elements, \emph{Physics of Fluids}, 8, 1169-1180.
\bibitem{Nikias1995} 
Nikias, C. L. and Shao, M. (1995). \emph{Signal Processing with $\alpha$-Stable Distributions and Applications}, John Wiley, New York.
\bibitem{Nolan1998} 
Nolan, J. P. (1998). Parameterizations and modes of stable distributions, \emph{Statistics and Probability Letters}, 38, 187-195.
\bibitem{Nolan2001} 
Nolan, J. P. (2001). Maximum likelihood estimation of stable parameters, In: Barndorff-Nielsen OE, Mikosch T, Resnick I, Eds. L\'{e}vy Processes: Theory and Applications. Boston: Birkh\"{o}user.
\bibitem{Gonzalez2009} 
Salas-Gonzalez, D., Kuruoglu, E. E., and Ruiz, D. P. (2009). Finite mixture of $\alpha$-stable distributions, \emph{Digital Signal Processing}, 19, 250-264. 
\bibitem{Gonzalez2010} 
Salas-Gonzalez, D., Kuruoglu, E. E., and Ruiz, D. P. (2010). Modelling with mixture of symmetric stable distributions using Gibbs sampling, \emph{Signal Processing}, 90, 774-783. 
\bibitem{Samorodnitsky1994} 
Samorodnitsky, G. and Taqqu, M. S. (1994). \emph{Stable Non-Gaussian Random Processes: Stochastic Models and Infinite Variance}, Chapman and Hall, London. 
\bibitem{Stapf1996} 
Stapf, S., Kimmich, R., Seitter, R.O., Maklakov, A.I., and Skid, V.D. (1996). Proton and deuteron field-cycling NMR relaxometry of liquids confined in porous glasses. \emph{Colloids and Surfaces: A Physicochemical and Engineering Aspects}, 115, 107-114
\bibitem{Teimouri2018}
Teimouri, M., Rezakhah, S., and Mohammdpour, A. (2018). EM algorithm for symmetric stable mixture model, \emph{Communications in Statistics-Simulation and Computation}, DOI: 10.1080/03610918.2017.1288244.
\bibitem{Uchaikin1999}
Uchaikin, V. V. and Zolotarev, V. M. (1999). \emph{Chance and Stability: Stable Distributions and Their Applications}, VSP, Utrecht.
\bibitem{Venables2002}
Venables, W. N. and Ripley, B. D. (2002). \emph{Modern Applied Statistics with S}, Fourth edition, Springer. 
\bibitem{Winterton1992}
Winterton, S. S., Smy, T. J., and Tarr, N. G. (1992). On the source of scatter in contact resistance data, \emph{Journal of Electronic Materials}, 21, 917-921.
\bibitem{Zolotarev1986}
Zolotarev, V. M. (1986). \emph{One-Dimensional Stable Distributions}, American Mathematical Society, Providence, R. I.
\end{thebibliography}
\end{document}